\documentclass[10pt,aps,pre,onecolumn,letterpaper,notitlepage,superscriptaddress,longbibliography]{revtex4-2}

\usepackage{amsmath}
\usepackage{amssymb}
\usepackage{bm}
\usepackage{tcolorbox}

\usepackage{hyperref}

\usepackage[utf8]{inputenc}

\newcommand{\kin}{k_\mathrm{in}}
\newcommand{\kout}{k_\mathrm{out}}

\newcommand{\uin}{u_\mathrm{in}}
\newcommand{\uout}{u_\mathrm{out}}

\newcommand{\G}{G}
\newcommand{\Gin}{{G_\mathrm{in}}}
\newcommand{\Gout}{{G_\mathrm{out}}}

\newcommand{\e}{\mathrm{e}}

\newcommand{\Pin}{{P_\mathrm{in}}}
\newcommand{\Pout}{{P_\mathrm{out}}}

\newcommand{\Qin}{{Q_\mathrm{in}}}
\newcommand{\Qout}{{Q_\mathrm{out}}}

\newcommand{\Fin}{{F_\mathrm{in}}}
\newcommand{\Fout}{{F_\mathrm{out}}}

\newcommand{\Sin}{\mathcal{S}}
\newcommand{\Sout}{\mathcal{P}}

\newcommand{\Var}{\mathrm{Var}}

\begin{document}

\author{Antoine Allard}
\affiliation{D\'epartement de physique, de g\'enie physique et d'optique, Universit\'e Laval, Qu\'ebec (Qu\'ebec), Canada G1V 0A6}
\affiliation{Centre interdisciplinaire en mod\'elisation math\'ematique, Universit\'e Laval, Qu\'ebec (Qu\'ebec), Canada G1V 0A6}
\affiliation{Vermont Complex Systems Center, University of Vermont, Burlington, VT 05405, USA}

\author{Cristopher Moore}
\affiliation{Santa Fe Institute, 1399 Hyde Park Road, Santa Fe, NM 87501, USA}

\author{Samuel V. Scarpino}
\affiliation{Pandemic Prevention Institute, Rockefeller Foundation, Washington, DC, USA}
\affiliation{Network Science Institute, Northeastern University, Boston, MA 02115, USA}
\affiliation{Santa Fe Institute, 1399 Hyde Park Road, Santa Fe, NM 87501, USA}
\affiliation{Vermont Complex Systems Center, University of Vermont, Burlington, VT 05405, USA}

\author{Benjamin M. Althouse}
\affiliation{Truveta, Inc., Bellevue, WA, USA}
\affiliation{University of Washington, Seattle, WA 98105, USA}
\affiliation{New Mexico State University, Las Cruces, NM 88003, USA}

\author{Laurent H\'{e}bert-Dufresne}
\affiliation{D\'epartement de physique, de g\'enie physique et d'optique, Universit\'e Laval, Qu\'ebec (Qu\'ebec), Canada G1V 0A6}
\affiliation{Department of Computer Science, University of Vermont, Burlington, VT 05405, USA}
\affiliation{Vermont Complex Systems Center, University of Vermont, Burlington, VT 05405, USA}

\title{The role of directionality, heterogeneity and correlations in epidemic risk and spread}

\begin{abstract}
  Most models of epidemic spread, including many designed specifically for COVID-19, implicitly assume mass-action contact patterns and undirected contact networks, meaning that the individuals most likely to spread the disease are also the most at risk to receive it from others.  Here, we review results from the theory of random directed graphs which show that many important quantities, including the reproduction number and the epidemic size, depend sensitively on the joint distribution of in- and out-degrees (``risk'' and ``spread''), including their heterogeneity and the correlation between them. By considering joint distributions of various kinds, we elucidate why some types of heterogeneity cause a deviation from the standard Kermack-McKendrick analysis of SIR models, i.e., so-called mass-action models where contacts are homogeneous and random, and some do not.  We also show that some structured SIR models informed by realistic complex contact patterns among types of individuals (age or activity) are simply mixtures of Poisson processes and tend not to deviate significantly from the simplest mass-action model.  Finally, we point out some possible policy implications of this directed structure, both for contact tracing strategy and for interventions designed to prevent superspreading events.  In particular, directed graphs have a forward and backward version of the classic ``friendship paradox''---forward edges tend to lead to individuals with high risk, while backward edges lead to individuals with high spread---such that a combination of both forward and backward contact tracing is necessary to find superspreading events and prevent future cascades of infection.
\end{abstract}

\maketitle

\section{Introduction}

In the field of network epidemiology, populations of hosts are modeled as graphs, where vertices are individuals and edges are contacts along which infectious diseases can spread~\citep{kiss2017mathematics}. Here, we review important concepts from the study of random graphs to inform our understanding of epidemics. We explain how these concepts and results can be used on the one hand to inform interventions and guide policy, and on the other hand to solve common nonlinear models of epidemic spread based on differential equations. Our paper is structured such that the main text focuses on concepts and insights, with all mathematical details presented in complementary boxes and in the Appendix. 

To model epidemic spreading with graphs, we consider general transmission graphs where we define edges as contacts or interactions that will transmit the disease from one individual to another should the first individual become infected. These transmission graphs can be generated through a set of assumptions~\citep{newman2002spread, meyers2007contact}, contact tracing data~\citep{hebert-dufresne2020r0}, or defined as isomorphic to classic epidemic models~\citep{kenah2007second}. We can then analyze how epidemic outcomes depend on the structure of these transmission graphs~\citep{kenah2007networkbased, meyers2007contact, kenah2011epidemic, miller2007epidemic}. The only assumption we make is that these graphs correspond to non-recurrent epidemic dynamics, i.e., where individuals have a state that represent their immunological status and never go back to previous states. This is true, for example, for infectious diseases that convey lifelong immunity. The Susceptible-Infectious-Recovered (SIR) model is the canonical example of such dynamics, but our discussion also applies to the Susceptible-Exposed-Infectious-Recovered (SEIR) model, or any other non-recurrent compartmental model of disease spread~\citep{anderson1992infectious}. The fact that individual states are non-recurrent here means that individuals can only be susceptible once and that every contact can only transmit the disease once, which is a key feature that allows us to analyze epidemic models based on the structure of the underlying graph.

Graphs allow us to model many kinds of heterogeneity and structure, and explore how and when heterogeneity affects epidemic thresholds and sizes.  For the simplest random graphs, epidemics end up being equivalent to classic mass-action compartmentalized models in epidemiology, which can be analyzed using systems of differential equations. As we will discuss, the SIR model on Erd\H{o}s-R\'enyi random graphs, where every pair of vertices is independently connected with the same probability, behaves like the Kermack–McKendrick SIR solution in the limit where the number of vertices---i.e. the population size---is infinitely large~\citep{kermack1927contribution, kermack1932contributions, kermack1933contributions, hebert-dufresne2020r0}.  Yet, since the \emph{degree} of a vertex is the number of edges it has, we can model superspreading events by having a heavy-tailed degree distribution, where some vertices have a degree considerably larger than the mean~\citep{lloyd-smith2005superspreading}. A random edge is more likely to be with a high-degree individual, who has more secondary contacts in turn. Thus the average degree of a vertex reached by a random edge is greater than the average degree of a random vertex. This phenomenon is often called the ``friendship paradox'': on average, your friends have more friends than you do~\citep{newman2018networks}. In the epidemic context, this phenomenon has two effects. First, the reproduction number is not simply the mean degree; it depends also on its second moment or variance (see Box~I). In particular, even if we hold the mean degree constant, increasing the variance lowers the critical transmission rate: the probability of transmission per contact needed to reach the epidemic threshold~\citep{newman2002spread}. Second, as developed in Refs.~\citep{muller2000contact, eames2003contact, kiss2006infectious} and recently pointed out as highly relevant for COVID-19 in Refs.~\citep{hellewell2020feasibility, kojaku2021effectiveness},  the ``friendship paradox'' can amplify the effectiveness of contact tracing, giving a better chance that a contact leads to a superspreading event, helping the prevention of further cascades of infection.

\begin{figure}[t]
    \centering
    \includegraphics{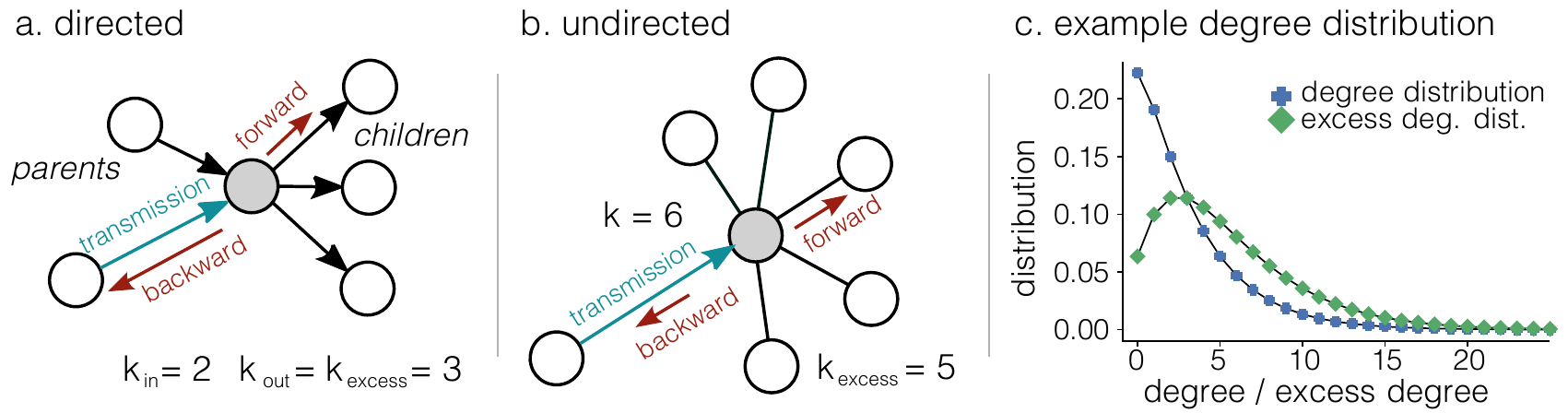}
    \caption{Definitions of key concepts around an index case.
    (\textbf{a}) In directed graphs, $\kin$ defines the number of potential parents of a vertex and $\kout$ its number of potential children.  Backward tracing goes up the genealogical tree toward predecessors while forward tracing goes down the tree toward descendants. Backward tracing will find a biased number of excess outgoing transmissions around the parent (backward friendship paradox, biased spread) while forward tracing would find a biased number of potential parents around the child (forward friendship paradox, biased risk).
    (\textbf{b}) In undirected graphs, transmissions can go both ways.  The identification of parent vertices for backward tracing is therefore only possible after a transmission event. The excess degree of a vertex will then refer to the number of edges other than the one which led to a transmission event and will be equal to the degree minus one (i.e. $k-1$).
    (\textbf{c}) Example of an undirected degree distribution (blue cross) along with the corresponding excess degree distribution (green diamond). The friendship paradox, i.e. a statistical sampling bias towards higher degree vertices, yields a more skewed tail and higher average in excess degrees than in original degrees.}
    \label{fig:excess_degree}
\end{figure}

\begin{tcolorbox}[title=Box I: Undirected graphs and the degree variance, floatplacement=b, float]
    \label{box:undirected}
    \small
    If $P(k)$ is the degree distribution in an undirected graph, i.e., the fraction of vertices with degree $k$, then the average \emph{excess degree}, or the expected number of secondary contacts resulting from a random contact [see Fig.~\ref{fig:excess_degree}(b)], is
    \begin{equation} \label{eq:r-undirected}
        R_0 = \frac{\sum_k k(k-1) P(k)}{\sum_k k P(k)}
            = \frac{\langle k^2 \rangle}{\langle k \rangle} - 1
            = \frac{\Var\,k}{\langle k \rangle} + \langle k \rangle - 1 \, .
    \end{equation}
    where $\langle k \rangle = \sum_k k P(k)$, $\langle k^2 \rangle = \sum_k k^2 P(k)$ and $\mathrm{Var}\,k = \langle k^2 \rangle - \langle k \rangle^2$.  An epidemic may only occur when $R_0 > 1$. For the Poisson degree distribution we have $\Var\,k = \langle k \rangle$ and thus $R_0=\langle k \rangle$, but for distributions with larger variance we have $R_0 > \langle k \rangle$. This is the friendship paradox: a random friend of yours has more friends, on average, than you do (even when not counting you).
\end{tcolorbox}

This view of superspreading events assumes that the transmission graph is undirected: that is, edges that would spread the disease in one direction would also spread it in the other.  While this may be true for individuals who attend a group event or work together in close quarters, some mechanisms of spread, such as surface contacts from handling and delivery, are inherently directed~\citep{meyers2006predicting}. In graph terms, an individual might have high out-degree but low in-degree, or vice versa. Intuitively, this is because they represent different concepts: in-degree captures the \textit{risk} that an individual receives the infection from others (network pathways for an incoming transmission event) while out-degree captures the \textit{spread} that an infected individual can cause once infected (the network pathways for outgoing transmission events). In addition, many non-pharmaceutical interventions to prevent disease spread have a directed effect. Wearing a simple mask is one such mechanism, as it is believed to reduce the probability that the wearer spreads a respiratory disease like COVID-19 to others, but not to strongly protect the wearer from receiving the disease from others \citep{desai2020masks, greenhalgh2020face}.

Most importantly, almost all epidemic models---including many mass-action models and models based on undirected contact networks---can be mapped unto epidemic percolation networks (EPNs)~\citep{kenah2007networkbased, kenah2007second, kenah2011epidemic} which are semi-directed (contain both directed and undirected edges).  A directed edge $a \to b$ in an EPN means that $a$ will transmit the disease to $b$ if $a$ becomes infected and $b$ is susceptible, and undirected edges simply correspond to two directed edges in opposite directions.  We therefore adopt a general view of directed transmission graphs~\citep{meyers2006predicting, bansal2006comparative, miller2007epidemic, kenah2007networkbased, miller2008bounding} which allows us to investigate the role of directionality, heterogeneity and correlations in these graphs without the need to specify the complex underlying mechanisms of a spreading dynamics, and without loss of generality.

\begin{figure}[t]
    \centering
    \includegraphics{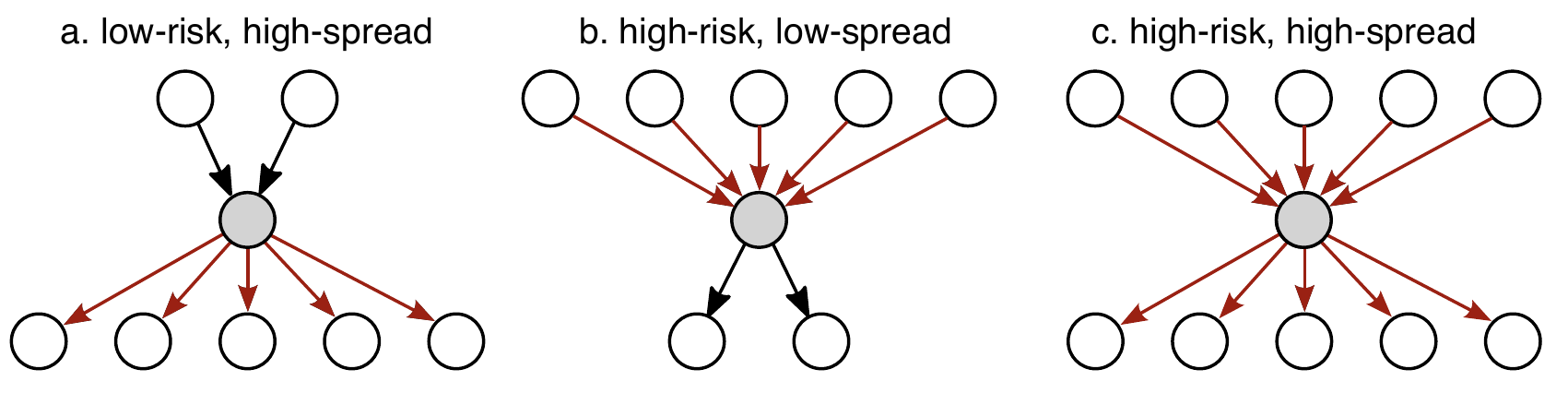}
    \caption{Different types of individuals in a directed graph model of disease spread: \textbf{a.} low-risk, but high-spread, \textbf{b.} high-risk, but low-spread, and \textbf{c.} high-risk and high-spread.}
    \label{fig:in-out}
\end{figure}

Each vertex in a directed graph has an \emph{in-degree}, i.e., the number of contacts from whom it could receive the disease, and an \emph{out-degree}, the number of contacts to whom it could spread the disease if it becomes infected. We can think of the in-degree as the \emph{risk} of this individual, and the out-degree as their \emph{spread}. As illustrated in Fig.~\ref{fig:in-out}, an individual might contribute to rapid spreading by having high risk, high spread, or both. For instance, an average immunocompromised individual might have high risk but low spread, while someone who fails to wear a mask even while those around them do might have low risk and high spread. Finally, someone who by choice or necessity interacts with many others in close quarters would have both high risk and high spread, as in the traditional undirected picture of superspreading events.

This distinction between risk and spread, or between incoming and outgoing edges, has several consequences. First, the reproduction number is proportional to the correlation between risk and spread (see Box~II). More precisely, the contribution of a given vertex to the reproduction number is proportional to the product of its in- and out-degrees. Vertices with both high risk and high spread are likely to become infected and then to infect many others. But if they have either high risk or low spread, then the expectation for the number of new cases they might generate is smaller. This has possible implications for non-pharmaceutical interventions: the effect of reducing an individual's spread is implicitly proportional to their risk.

\begin{tcolorbox}[title=Box II: Directed graphs and correlation between risk and spread,floatplacement=t,float,label=box:directed]
    \small
    In a directed graph, each vertex has an \emph{in-degree} $\kin$ and an \emph{out-degree} $\kout$.  We associate these respectively with their \emph{risk}, the number of contacts from whom they could receive the infection, and their \emph{spread}, the number of contacts to whom they will transmit it, if they become infected.  The degree distribution is then a joint distribution, giving the fraction $P(i,j)$ of vertices with $\kin=i$ and $\kout=j$. Since the total number of incoming and outgoing edges must match, the mean in-degree and the mean out-degree are equal:
    \begin{equation} \label{eq:in-out-equal}
        \langle \kin \rangle
            = \sum_{i,j} i P(i,j)
            = \langle \kout \rangle
            = \sum_{i,j} j P(i,j) \, .
    \end{equation}
    The \emph{directed configuration model} of random graphs~\citep{dorogovtsev2001giant, boguna2005generalized, meyers2006predicting} assumes that vertices are connected to each other randomly conditioned on their in- and out-degrees. The ``heads'' of the directed edges are randomly matched with their ``tails,'' and each outgoing edge arrives at a given vertex with probability proportional to that vertex's in-degree. This method of constructing a random graph can give rise to self-loops or parallel edges, but these are rare in sparse graphs (with finite average degree). In the limit of large population size, we can also say that for each pair of vertices $a,b$ there is an edge $a \to b$ with probability $\kout(a) \kin(b) / m$ where $m=\langle \kin \rangle n$ is the total number of edges and $n$ is the total number of vertices.\\

    The reproduction number, or the expected number of secondary transmissions resulting from a random transmission, is
    \begin{equation} \label{eq:r-directed}
        R_0 = \frac{\sum_{i,j} ij P(i,j)}{\sum_{i,j} i P(i,j)}
            = \frac{\langle \kin \kout \rangle}{\langle \kin \rangle} \,.
    \end{equation}
    Just as $R_0$ depends on the variance of the degree distribution in the undirected case, it now depends on the covariance, or correlation, between the in- and out-degrees. In particular, the contribution of each individual to the reproduction number is proportional to the product of their risk and their spread.
\end{tcolorbox}

Second, we now have two distinct types of friendship paradox, a forward one and a backward one (see Fig.~\ref{fig:excess_degree} and Box~III). Outgoing edges connect to vertices with probability proportional to their in-degree, so following an edge forward will tend to lead to individuals with high risk. Incoming edges, in contrast, come from vertices with probability proportional to their out-degree, so going following edges backward tends to lead to individuals with high spread.

This suggests that to find superspreading events, and thus to prevent a large number of downstream cases, it may be important to perform \emph{backward} contract tracing, determining the source of an index case, rather than focusing only on \emph{forward} contact tracing, finding those who the index case may have infected.  As shown in Box~III, the average spread of the predecessor of an index case is larger than that of an average successor, especially if the spread has high variance, except in the unusual case where risk and spread are almost perfectly correlated. This insight also implies that the time window we use to define relevant contacts should include their likely predecessor as well as the likely successors of an index case.  We can then trace the ``siblings'' of the index case, i.e., others who were infected by the same individual, and prevent a larger number of future cases.

\section{Epidemic Probability and Epidemic Size\label{sec:epidemic}}

Here we review some results of Refs.~\citep{newman2001random, cooper2004size} from random graph theory which show how to calculate the epidemic size as a function of the joint distribution of individual risk and spread. We note that these results can be generalized~\citep{dorogovtsev2001giant, timar2017mapping} and expanded to include semi-directed graphs composed of a mixture of both directed and undirected edges~\citep{boguna2005generalized, meyers2006predicting}.

As mentioned above, we define the graph such that an edge $a \to b$ means that $a$ will transmit the disease to $b$ if $a$ becomes infected and $b$ is susceptible.  These transmission graphs are called epidemic percolation networks (EPNs)~\citep{kenah2007networkbased, kenah2007second, kenah2011epidemic}. The term ``EPN'' is often specifically used to refer to transmission graphs whose structure is isomorphic to specific epidemic dynamics and are therefore typically semi-directed.  In other words, EPNs consist in randomly chosen subgraphs of a more general graph of contacts that have the potential to transmit the disease, and in which each transmission occurs with some probability.  Here we focus on general but purely directed graphs for the sake of simplicity, without affecting the generality of our conclusions (the phenomenology observed below remains the same in the presence of undirected edges).  Our approach allows us to consider general forms of heterogeneities and correlations without the need to specify an underlying epidemic dynamics.  The connection between general contact networks and EPNs is discussed in Box~V.

Let us first compute the epidemic probability, i.e., the probability that a single infected individual generates a large-scale epidemic as opposed to a small outbreak. It helps to think of a process where edges, rather than vertices, propagate the disease from one generation to the next. An outgoing edge produces $j$ ``children,'' or secondary contacts, if the edge arrives at a vertex with out-degree $j$. A central question in the theory of branching processes is whether the resulting downstream tree of contacts is finite or infinite.  Since the total number of descendants of a vertex is finite if and only if this is true of all of its children, the probability of this event obeys a fixed-point equation which can be written compactly using probability generating functions (see Box~IV).

Let us now consider the epidemic size. Vertices can receive the infection from one of their incoming edges, or predecessors, who received it from their predecessor in turn.  We can thus consider a backwards branching process of incoming edges, where an edge has $i$ ``parents'' (predecessors upstream in the directed graph) if it comes from a vertex with in-degree $i$. If a vertex has only a finite number of ancestors, counting all upstream vertices, then with high probability none of them will be the index case and it will remain uninfected. But if it has a very large (i.e. infinite) number of ancestors, then with high probability one of its ancestors will be the index case and the vertex will eventually become infected. Thus the calculation of the epidemic size is equivalent to the calculation of the epidemic probability in a time-reversed model, where the disease spreads backward from each vertex to its predecessors.

On undirected graphs, these forward and backward branching processes can often be the same, such that the probability that they die out obeys the same fixed-point equation.  This leads to the observation that, in very simple percolation processes, the epidemic size equals the epidemic probability.  As explained previously, more complex and realistic processes imply the presence of directed edges, meaning that these two quantities are typically different~\citep{newman2002spread, boguna2005generalized, meyers2006predicting}.  Therefore, in the classic SIR dynamics, these fixed-point equations are expected to be different since the EPNs capturing this process are semi-directed (see Box~V for details); with the epidemic probability being smaller than the relative size of the epidemic as predicted by the Kermack-McKendrick analysis~\citep{kenah2007second, kuulasmaa1982spatial}. In the next section, we consider degree distributions of various kinds and with varying levels of correlation between risk and spread, and look at how these types of heterogeneity affect their epidemic size.

\begin{tcolorbox}[title=Box III: Forward and backward friendship paradoxes and contact tracing, floatplacement=t, float, label=box:forward-backward]
    \small
    Suppose we are performing contact tracing from an index case.  Since the probability that a given individual is the predecessor of the index case (i.e. the source of their infection) is proportional to the out-degree (or spread) of the former, the average excess spread of their predecessor---that is, the average number of \textit{other} children the predecessor has---is
    \begin{equation} \label{eq:spread-pred}
        \langle \kout \rangle_\textrm{predecessor}
            = \frac{\sum_{i,j} j(j-1) P(i,j)}{\sum_{i,j} j P(i,j)}
            = \frac{\langle \kout^2 \rangle}{\langle \kin \rangle} - 1
            = \frac{\Var\,\kout}{\langle \kin \rangle} + \langle \kin \rangle - 1 \, .
    \end{equation}
    Similarly, the probability a given individual is a successor of the index case is proportional to their in-degree or risk, so the average excess risk of a successor---that is, the average number of \textit{other} predecessors the successor had---is
    \begin{equation} \label{eq:risk-succ}
        \langle \kin \rangle_\textrm{successor}
            = \frac{\sum_{i,j} i(i-1) P(i,j)}{\sum_{i,j} i P(i,j)}
            = \frac{\langle \kin^2 \rangle}{\langle \kin \rangle} - 1
            = \frac{\Var\,\kin}{\langle \kin \rangle} + \langle \kin \rangle - 1 \, .
    \end{equation}
    These are the backward and forward friendship paradoxes: the predecessor of a random vertex tends to have high spread, but a successor tends to have high risk, especially when both of these have high variance. On the other hand, the average spread of a successor is just the reproduction number calculated in Eq.~\eqref{eq:r-directed} above,
    \begin{equation} \label{eq:spread-succ}
        \langle \kout \rangle_\textrm{successor}
            = R_0
            = \frac{\langle \kin \kout \rangle}{\langle \kin \rangle} \, .
    \end{equation}
    Comparing this with Eq.~\eqref{eq:spread-pred}, we see that if the spread $\kout$ has high variance then the predecessor has larger spread on average than a successor, unless the risk and spread are almost perfectly correlated, i.e., unless $\langle \kin \kout \rangle \ge \langle \kout^2 \rangle-1$. This insight highlights the conclusion of Ref.~\citep{kojaku2021effectiveness} that contact tracing strategies should include tracing backwards from the index case as well as forwards. In fact, this recommendation becomes critical once we consider the directionality of contacts such that risk and spread might not be perfectly correlated.
\end{tcolorbox}
\clearpage

\begin{tcolorbox}[title=Box IV: Generating functions formalism for epidemic size and probability,floatplacement=t,float,label=box:epidemic-size]
    \small
    Given the degree distribution $P(i,j)$, we define the following probability generating function (PGF)
    \begin{equation} \label{eq:G}
        \G(x, y) = \sum_{i,j=0}^\infty P(i,j) x^i y^j \, .
    \end{equation}
    Many quantities can be written compactly in terms of $G$ and its derivatives, including the average in- and out-degrees, Eq.~\eqref{eq:in-out-equal}, and the reproduction number, Eq.~\eqref{eq:r-directed}:
    \begin{equation} \label{eq:g-derivs}
        \langle \kin \rangle 
            = \left. \frac{\partial G(x, y)}{\partial x} \right\vert_{x=y=1}
            = \langle \kout \rangle
            = \left. \frac{\partial G(x, y)}{\partial y} \right\vert_{x=y=1} \, ,
        \quad 
        R_0 = \frac{1}{\langle \kin \rangle} \left. \frac{\partial^2 G(x, y)}{\partial x \,\partial y} \right\vert_{x=y=1}
            = \frac{\langle \kin \kout \rangle}{\langle \kin \rangle} \, . 
    \end{equation}
    We can also use $G$ to analyze branching processes on the graph. In the forward branching process of an outbreak, the probability $\Qout(j)$ that a random contact results in $j$ secondary contacts, i.e., that a random edge arrives at a vertex of out-degree $j$, is 
    \begin{equation} \label{eq:qout}
        \Qout(j) 
            = \frac{\sum_i i P(i,j)}{\sum_{i,j} i P(i,j)} 
            = \frac{1}{\langle \kin \rangle} \sum_i i P(i,j) \, . 
    \end{equation}
    We can write a generating function for $\Qout$ in terms of $G$:
    \begin{equation} \label{eq:fout}
        \Fout(y) 
            = \sum_j \Qout(j) y^j 
            = \frac{1}{\langle \kin \rangle} \left. \frac{\partial G(x,y)}{\partial x} \right\vert_{x=1}
    \end{equation}
    Now let $\uout$ be the probability that an edge has a finite number of descendants, i.e. that it will lead to a finite outbreak rather than an epidemic. This is true if it is true for all $j$ of its children. Since these are independent, averaging over $j$ gives the fixed-point equation
    \begin{equation} \label{eq:uout}
        \uout 
            = \sum_j \Qout(j) \uout^j 
            = \Fout(\uout) \, ,
    \end{equation}
    of which $\uout$ is the smallest non-negative solution. Finally, since a vertex with out-degree $j$ generates an epidemic with probability $1-\uout^j$, the probability that infecting a random initial vertex leads to an epidemic is
    \begin{equation} \label{eq:epidemic-prob}
        \Sout
            \equiv 1 - \sum_{i,j} P(i,j) \uout^j 
            = 1 - G(1,\uout) \, .
    \end{equation}
    To compute the epidemic size, we consider the backwards branching process of incoming edges. The probability $\Qin(i)$ that an edge comes from a vertex with in-degree $i$, and the corresponding generating function, is
    \begin{equation} \label{eq:qin}
        \Qin(i) 
            = \frac{1}{\langle \kin \rangle} \sum_j j P(i,j) 
        \quad \text{and} \quad
        \Fin(x) 
            = \sum_i \Qin(i) x^i 
            = \frac{1}{\langle \kin \rangle} \left. \frac{\partial G(x,y)}{\partial y} \right\vert_{y=1} \, . 
    \end{equation}
    The probability $\uin$ that an edge has a finite number of ancestors is then the smallest non-negative solution of
    \begin{equation} \label{eq:uin}
        \uin = \sum_i \Qin(i) \uin^i 
             = \Fin(\uin) \, .
    \end{equation}
    Finally, a vertex with in-degree $i$ becomes infected with probability $1-\uin^i$, so the size of the epidemic is
    \begin{equation} \label{eq:epidemic-size}
        \Sin 
            \equiv 1 - \sum_{i,j} P(i,j) \uin^i 
            = 1 - G(\uin,1) \, .
    \end{equation}
    Note that Eqs.~\eqref{eq:uout}~and~\eqref{eq:uin} assume that $\Qout$ and $\Qin$ stay constant as the disease spread on the graph, and that the states of the neighbors of a given vertex with respect to them being part of the epidemic are not correlated~\cite{newman2001random, newman2002spread}.  These assumptions imply that Eqs.~\eqref{eq:epidemic-prob}~and~\eqref{eq:epidemic-size} are only exact in the limit of infinitely large graphs.
\end{tcolorbox}
\clearpage

\begin{figure}
    \centering
    \includegraphics{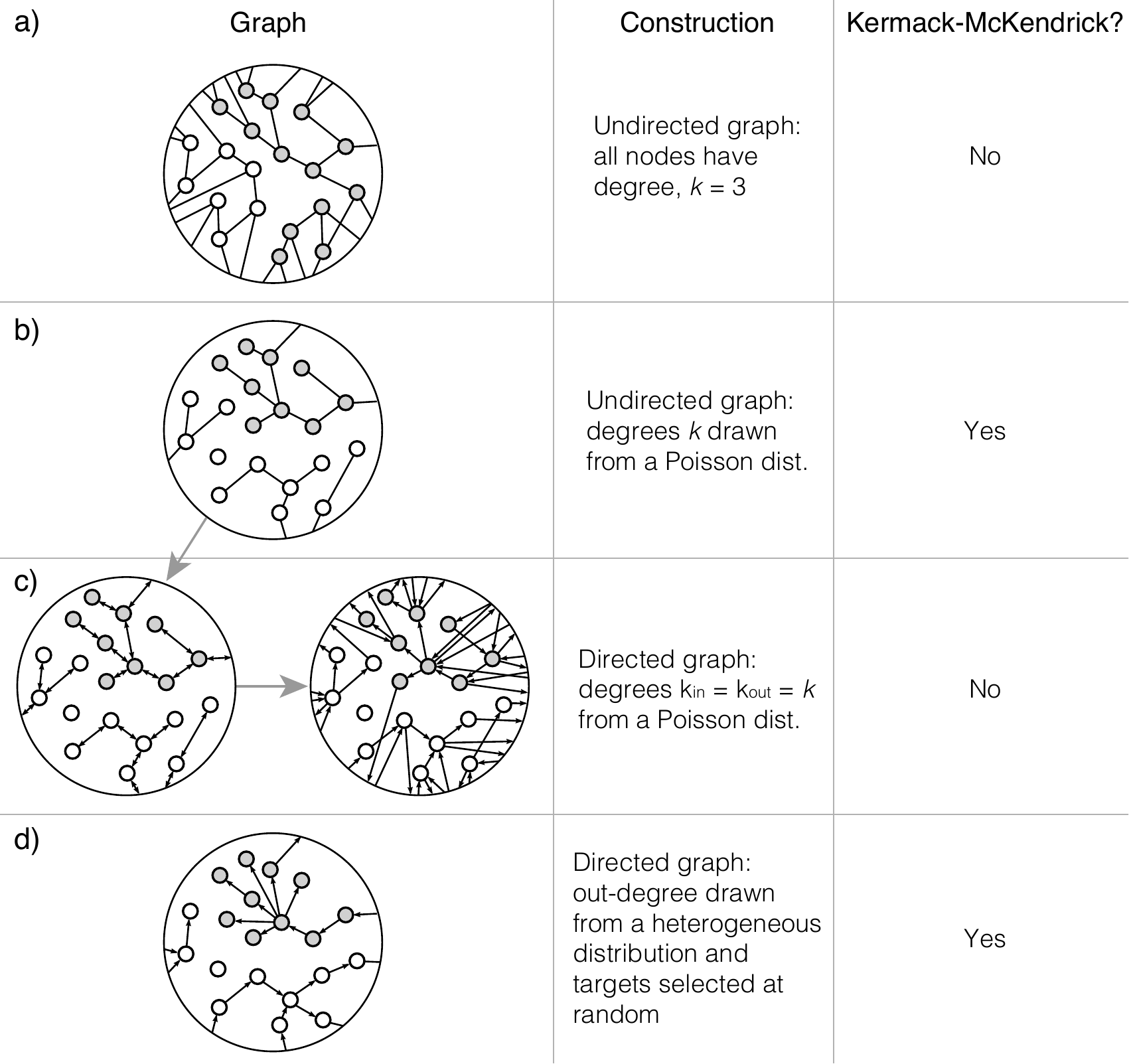}
    \caption{Four examples of transmission graphs, their construction, and whether the Kermack-McKendrick solution corresponds to the expected epidemic size. \textbf{(a)} An undirected graph where all vertices are exactly the same, showing that too much homogeneity is different from mass-action mixing. \textbf{(b)} The simple Poisson graph, the fish out of water, is the only type of undirected transmission graph where the Kermack-McKendrick solution holds. Simple Poisson graphs can be represented with a network structure the expected epidemic size corresponds exactly on that of the classic mass-action models and therefore fail to include important network effects. \textbf{(c)} We take the undirected Poisson graph and add directions to all edges (left). This procedure preserves the in- and out-degree for all vertices. We then rewire the directed edges to obtain a simple graph without overlap between contacts (right). Should we expect the Kermack-McKendrick solution to correctly predict the epidemic size as it did in the original undirected graph? No. \textbf{(d)} A directed graph obtained by independently specifying an in-degree distribution (here Poisson) and an out-degree distribution (here fairly heterogeneous). This corresponds to a situation where all individuals have similar risk of infection, while allowing for superspreading events. Proofs of mapping (or lack thereof) to the Kermack-McKendrick solution are provided in Appendix.}
    \label{fig:examples}
\end{figure}
\clearpage

\begin{figure}[h!!!]
    \centering
    \includegraphics{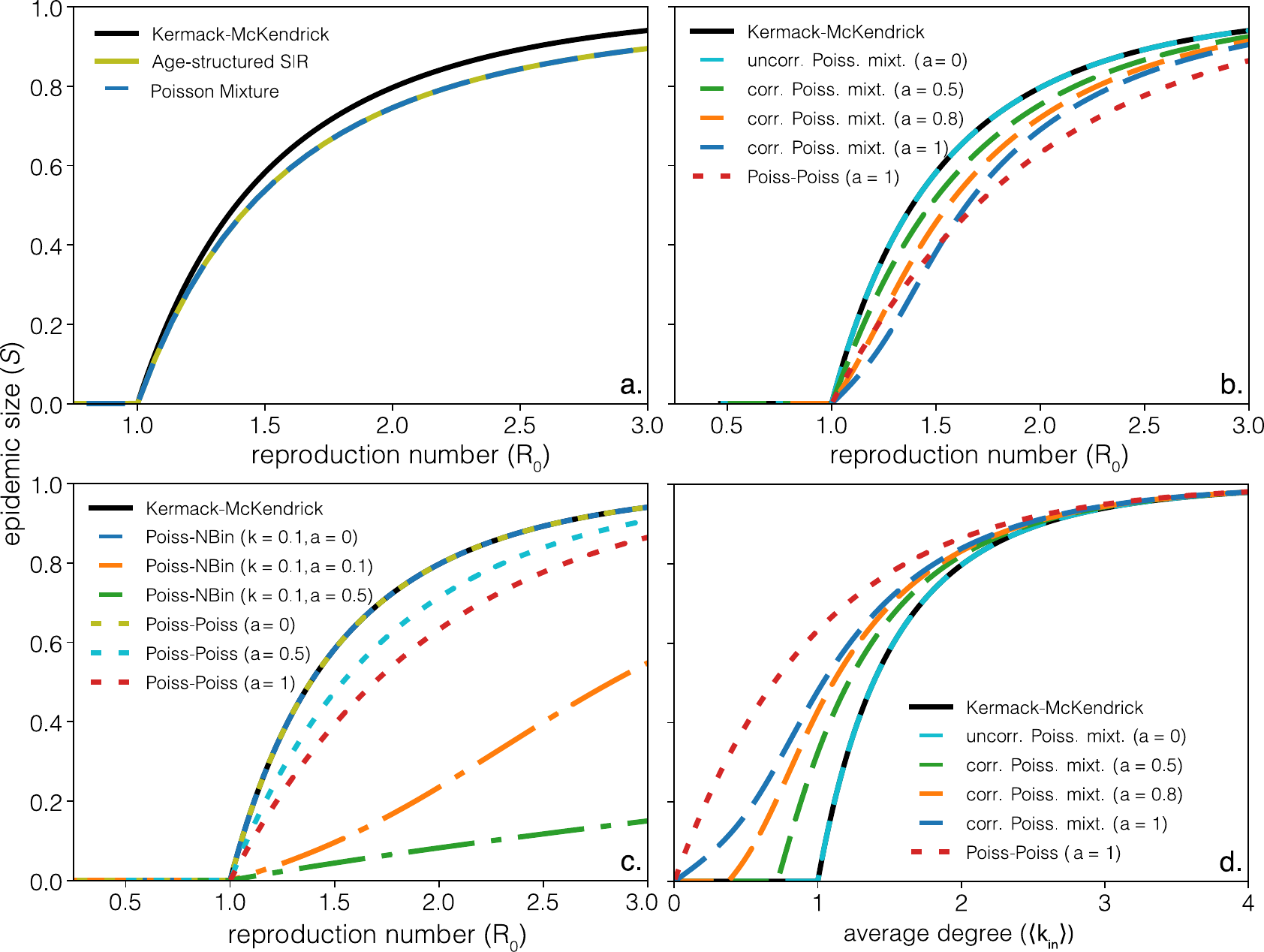}
    \caption{
        (\textbf{a}) Comparison between Age-structured SIR and Poisson mixture models.  The Poisson mixture model presented in Box~VI is compared to the numerical integration of the model of Ref.~\citep{zhang2020changes} given by Eqs.~\eqref{eq:SIR} and to the Kermack-McKendrick solution given by Eq.~\eqref{eq:KMK_solution}.  This comparison showcases the versatility of our approach, and illustrates how many mass-action models similar to Eqs.~\eqref{eq:SIR} are equivalent to transmission graphs where in-degrees and out-degrees are a mixture of Poisson distributions when it comes to predicting the epidemic size.
        (\textbf{b}) Impact of correlations on epidemic size. We consider correlated Poisson mixture model (see Sec.~\ref{sec:corr_poiss_mixt_model} of the Appendix) with $w_0=w_1=0.5$ and $z_{11}^\mathrm{in} = z_{12}^\mathrm{in} = z_{21}^\mathrm{in} = z_{22}^\mathrm{in}$.  The parameter $a$ controls the correlation between the in-degrees and out-degrees for vertices of type 1: $a=0$ corresponds to no correlation while $a=1$ corresponds to perfect correlation.  The case $a=0$ is equivalent to the uncorrelated scenario explored in Sec.~\ref{sec:kmk_derivation} of the Appendix, while the dotted red line corresponds to the fully correlated scenario explored in Sec.~\ref{sec:full_correlation} of the Appendix.
        (\textbf{c}) Impact of correlations and heterogeneity on the final epidemic size.  In each case study, the in-degree is distributed according to a Poisson [$\Gin(x) = \mathrm{e}^{R_0(x-1)}$] and the out-degree is either distributed according to a Poisson [$\Gout(y) = \mathrm{e}^{R_0(y-1)}$] or a negative binomial distribution [$\Gout= (1 + R_0 (1 - y)/k)^{-k}$].  The parameter $a$ controls the correlation between the in-degrees and out-degrees: $a=0$ corresponds to no correlation while $a=1$ corresponds to perfect correlation, see Eq.~\eqref{eq:general_G} for details.  The epidemic size, $\mathcal{S}$, and the reproduction number, $R_0$, were computed with Eqs.~\eqref{eq:uin_correlations}--\eqref{eq:R0_correlation}.  The Kermack-McKendrick solution [Eq.~\eqref{eq:KMK_solution}] was added to illustrate the impact of heterogeneity and correlations.  Importantly, we show that increasing correlations always lead to lower final epidemic size, and that their effect is larger for higher heterogeneity in out-degree.
        (\textbf{d}) Same results as shown in panel (b) but plotted against the average degree of the populations rather than $R_0$. This emphasizes multiple results. First, correlations lower the epidemic threshold. Second, correlations in a subset of the population can lead to sequential epidemic transitions~\cite{allard2017asymmetric,hebert-dufresne2019smeared}, here visible as an S-shaped curve for the cases $a=0.8$ and $a=1.0$. Third, differences between the Kermack-McKendrick and correlated transmission graphs shown in panel (a) are partially hidden in the $R_0$ calculation.
    }
    \label{fig:SIR}
\end{figure}
\clearpage

\begin{tcolorbox}[title=Box V: Probabilistic transmission and epidemic percolation networks,label=box:EPN]
    \small
    There is a long shared history between epidemic models, percolation processes, and contact networks~\citep{frisch1963percolation, dietz1967epidemics, mollison1977spatial, kuulasmaa1982spatial, grassberger1983critical,cardy1985epidemic,sander2002percolation}.
    In this literature, there are two general approaches. The first and simplest assumes (or makes the approximation) that epidemics spread as a bond percolation process: transmissions on contacts between infectious and susceptible individuals are independent stochastic events \cite{newman2002spread, meyers2005network, meyers2006predicting, meyers2007contact, funk2010interacting, dimitrov2010mathematical}. These probabilities can be heterogeneous and directed to account for diverse types of contacts and individuals \cite{allard2009heterogeneous, allard2015general, allard2017asymmetric, artime2021percolation} but always assume that all contacts of an infectious vertex transmit independently. However, in most epidemic models, these transmissions events are correlated through the recovery period of the infectious vertex. The second approach therefore defines a different percolation process designed to be isomorphic to classic, dynamical, epidemic models \cite{kenah2007second, kenah2007networkbased,miller2007epidemic}.  In this work, we have considered transmission trees stemming from both approaches.

    \textit{Bond percolation as an epidemic model}---In general, the entire PGF formalism introduced in the main text can be applied to the study of a percolation process. Recall that the variables  $x$ and $y$ are simply counting variables. If a contact only exist in the transmission tree with a probability $q$ corresponding to the transmission probability, then it contributes $qx+(1-q)$ to the counting function (counting $x$ only with probability $q$). We showed that transmission trees led to an epidemic size agreeing with the Kermack-McKendrick solution if the in-degree is Poisson-distributed with mean $R_0 = \langle \kin \rangle$, generated with
    \begin{equation} \label{eq:g-poisson}
        \Gin(x)
            = \sum_{i=0}^\infty P(i) x^i
            = \sum_{i=0}^\infty \frac{\e^{-\langle \kin \rangle} \langle \kin \rangle^i}{i!} \,x^i
            = \e^{\langle \kin \rangle (x-1)} \, ,
    \end{equation}
    The bond percolation model leads to the same scenario if, for instance, we consider the case where each individual has $d$ contacts, each transmitting the disease with probability $q$. Then $\langle \kin \rangle = q d$ and the distribution of the in-degrees follows
    \begin{equation} \label{eq:pgf-binomial}
        \Gin(x)
            = \sum_{i=0}^d P(i) x^i
            = \sum_{i=0}^d \binom{d}{i} q^i (1-q)^{d-i} x^i
            = \big( qx + (1-q) \big)^{d}
            = \left( 1 + \frac{\langle \kin \rangle (x-1)}{d} \right)^{\!d} \, .
    \end{equation}
    For large degree $d$, the identity $(1+w/d)^d \approx \e^{w}$ applies, and Eq.~\eqref{eq:pgf-binomial} falls back on Eq.~\eqref{eq:g-poisson}, hence recovering the Kermack-McKendrick prediction for epidemic size.

    \textit{Epidemic percolation networks as isomorphism of epidemic models}---To account for the fact that transmissions are correlated through the recovery period of a vertex, we must add heterogeneity in the percolation process. To model the final size of the SIR model with transmission rate $\beta$ and recovery rate $\gamma$, epidemic percolation networks can be generated like so
    \begin{enumerate}
        \item For all vertices $i$ in the contact network, assign a recovery period $\tau_i$, drawn from $\gamma e^{-\gamma\tau_i}$ based on the Poisson recovery process of the SIR model.
        \item For each ordered pair of contacts between vertices (say $i$ and $j$), draw a directed edge from $i$ to $j$ with probability $T_{ij}(\tau_i) = (1-e^{-\beta\tau_i})$ based on the Poisson transmission process over infection time $\tau_i$ of the SIR model.
    \end{enumerate}
    These epidemic percolation networks can then be analyzed with the PGF formalism described in the main text. The process can be straightforwardly generalized to other epidemic dynamics by calculating the correct distributions for $\tau_i$ and $T_{ij}(\tau_i)$ \cite{kenah2007second}. Importantly, this process breaks the symmetry between epidemic probability and size of the bond percolation approach, yielding smaller epidemic probabilities~\cite{kenah2007second}, and reduces to classic bond percolation only when $\tau_i = \tau$ for all vertices~\cite{kuulasmaa1982spatial, kenah2007second}.
\end{tcolorbox}
\clearpage

\section{Examples\label{sec:examples}}

Let us first consider a simple and random undirected transmission graph, as illustrated in Fig.~\ref{fig:examples}(a-b) whose structure is fully specified by its degree distribution, $P(k)$, setting the number of contacts, $k$, leading to and from a given vertex.  Since the graph is undirected, we have $\kin = \kout = k$.  One could obtain such a transmission graph by either directly assuming undirected transmission mechanisms such as a percolation process~\citep{cardy1985epidemic} or by starting from epidemic dynamics (e.g., the SIR model) and applying a set of approximations~\citep{newman2002spread}. In the context of EPNs, undirected transmission graphs are possible for simple SI dynamics, or in very specific cases where transmission of a disease is certain over some type of contacts and impossible otherwise.  As shown in Ref.~\citep{hebert-dufresne2020r0}, the epidemic size $\mathcal{S}$ obtained on these graphs corresponds to the classic Kermack-McKendrick solution, meaning that it is the smallest non-negative solution of 
\begin{equation} \label{eq:KMK_solution}
    \mathcal{S} = 1 - \mathrm{e}^{-R_0\mathcal{S}} \ ,
\end{equation}
if and only if the degree distribution is Poisson, i.e., $P(k) = R_0^k e^{-R_0} / k!$ where $R_0$ is the reproduction number.  What is so special about this distribution? As it turns out, the Poisson distribution is the only one where the excess degree has the same distribution as the degree itself: the probability that an edge leads to $k$ secondary transmissions equals the probability that a random vertex does the same.  In terms of the friendship paradox, the probability that you have $k$ friends is the same as the probability that your friend has $k$ friends in addition to you. In particular, these two distributions have the same mean, so $R_0 = \langle k \rangle$.  This is similar to the mass-action hypothesis of Kermack and McKendrick as it implies that information about the contact structure does not matter.  For any other degree distribution---even a more homogeneous one as in Fig.~\ref{fig:examples}(a)---the distribution of excess degrees is different from the degree distribution, and the Kermack-McKendrick solution given by Eq.~\eqref{eq:KMK_solution} for the epidemic size does not apply.

Consider now a directed version of the Poisson graph. As shown in Fig.~\ref{fig:examples}(c), we can think of undirected edges as bidirectional contacts, i.e. pairs of directed edges, and make this directionality explicit before randomizing the incoming and outgoing edges. What we now have is a graph where in-degrees and out-degrees are equal ($\kin = \kout = k$) and are still drawn from a Poisson distribution, but where incoming and outgoing neighbors are different.  Even if the construction of the directed graph structure is very close to that of the undirected case, this directed version does not fall back on the Kermack-McKendrick solution given by Eq.~\eqref{eq:KMK_solution}~(see Sec.~\ref{sec:full_correlation} of the Appendix).  One key difference can be observed around transmission dead-ends, i.e., vertices with $\kout = 0$, which are now impossible to reach since they also have $\kin = 0$.  In other words, if a vertex has been reached via one if its incoming edges, there is always at least one outgoing edge available to reach the next vertex.  As shown in Sec.~\ref{sec:kmk_derivation} of the Appendix, equivalence with the Kermack-McKendrick solution is only recovered when in-degrees and out-degrees are uncorrelated.

It turns out that some naive approaches to modeling heterogeneity do not, in fact, alter the epidemic size at all. 
Consider the following kind of simulation, which would be run until there are no newly infected individuals and where $\Pout$ is an arbitrary distribution of out-degrees:
\begin{enumerate}
    \item Start with a single new infectious case in an otherwise susceptible population of size $N$.
    \item For each newly infected individual $i$, draw a number $\kout$ of transmissions from the out-degree distribution $\Pout$.
    \item Randomly select $\kout$ individuals from the population. For each one, if it is susceptible then mark it as newly infected and go to step 2, otherwise do nothing.
\end{enumerate}
This type of simulation generates transmission trees similar to the example shown in Fig.~\ref{fig:examples}(d). While the out-degree distribution $\Pout$ can be arbitrarily heterogeneous, including various kinds of superspreading events, the incoming edges fall with equal probability on every susceptible individual because of the random selection of neighbors at step 3. As a result, the distribution of risk, or in-degree, is Poisson, and the in- and out-degrees are uncorrelated.  As shown in Sec.~\ref{sec:kmk_derivation} of the Appendix, in the limit of large $N$, the typical epidemic size produced by this kind of simulation exactly corresponds to the Kermack-McKendrick solution regardless of the out-degree distribution $\Pout$.

These simple examples suggest that the Kermack-McKendrick solution given by Eq.~\eqref{eq:KMK_solution} relies on (i) a Poisson distribution of risk among individuals and (ii) a lack of correlation between risk and spread.  It does not actually rely on homogeneous spread but only on the independence between the infection risk faced by an individual and the number of secondary infections caused by that individual if infected. The undirected Poisson case, where this independence does not hold since the in- and out-degrees are equal by definition, is somewhat unique since the lack of backtracking and the friendship paradox balance out. We formalize this intuition mathematically in the Appendix, and some important results are visually summarized in Fig.~\ref{fig:SIR}.

\begin{tcolorbox}[title=Box VI: Poisson mixture model, floatplacement=t, float]
    \label{box:poisson-mixture}\small
    \renewcommand{\Fin}[1]{{F_{#1}^\mathrm{in}}}
    \renewcommand{\uin}[1]{{u_{#1}^\mathrm{in}}}
    \small
    We consider a general case in which there are $M$ types of vertices and where a fraction $w_\ell$ of the vertices are of type $\ell$.  We assume that the in-degrees and the out-degrees are uncorrelated, and that the contribution to the in-degree from each vertex type is independent and distributed according to a Poisson distribution.  Thus, a PGF $\G_\ell$ is defined for each vertex type and has the general form
    \begin{align}
        G_\ell(\bm{x},\bm{y}) = \mathrm{exp}\left[ \sum_{r=1}^M z_{\ell r}^\mathrm{in} (x_{r} - 1) \right] \Gout(\bm{y}) \ ,
    \end{align}
    where $z_{\ell r}^\mathrm{in}$ is the average number of incoming edges type-$\ell$ vertices received from vertices of type $r$, and where $\Gout(\bm{y})$ is an arbitrary PGF. The average in-degree of type-$\ell$ vertices is therefore $\sum_r z_{\ell r}^\mathrm{in}$.  Akin to Eq.~\eqref{eq:in-out-equal}, the average in- and out-degrees between any pair of vertex types are related via
    \begin{align}
        w_i z_{ij}^\mathrm{out} = w_j z_{ji}^\mathrm{in} \, ,
    \end{align}
    which can be seen as the conservation of the total number of contacts between types of vertices (i.e. there are as many edges ``leaving type $i$ vertices towards type $j$ vertices'' as there are edges ``arriving at type $j$ from type $i$ vertices'').  Note that the matrices whose elements are respectively $\{z_{r\ell}^\mathrm{in}\}$ and $\{z_{r\ell}^\mathrm{out}\}$ need not be symmetric nor satisfy $w_i z_{ij}^\mathrm{in} = w_j z_{ji}^\mathrm{in}$ and $w_i z_{ij}^\mathrm{out} = w_j z_{ji}^\mathrm{out}$.  Following Ref.~\citep{allard2015general}, the epidemic size is 
    \begin{align}
        \Sin = 1 - \sum_{\ell=1}^{M} w_\ell G_\ell(\bm{\uin{}}, \bm{1})
    \end{align}
    where $\bm{\uin{}} = (\uin{1},\ldots,\uin{M})$ is the smallest non-negative solution of
    \begin{align}
      \uin{\ell} = \mathrm{exp}\left[ \sum_{r=1}^M z_{\ell r}^\mathrm{in} (\uin{r} - 1) \right]
    \end{align}
    for $\ell=1,\ldots,M$.  The reproduction number, $R_0 = \rho(\mathbf{Z})$, is equal to the spectral radius, i.e., the largest absolute eigenvalue, of the matrix $\mathbf{Z}$ whose elements are $\{z_{\ell r}^\mathrm{in}\}$.
\end{tcolorbox}

\section{Connection with structured differential equation models\label{sec:sir}}
The mathematical analysis presented thus far is completely general and exact on very large (formally infinite) random directed graphs.
To further assess the impact of risk heterogeneity on the epidemic size, we need to look into the mechanisms behind the joint distribution of both risk and spread in specific models.  Here we show how one case of these distributions arises from age-structured models in epidemiology.

One particularly powerful and common approach in current models of the COVID-19 pandemic is age-structured SIR or SEIR models (e.g., Refs.~\citep{fumanelli2012inferring, mistry2021inferring, zhang2020changes}) where individuals are separated in different types based on their age and where contacts between age groups are specified through a contact (or ``mixing'') matrix. This approach is powerful because it allows the inclusion of important sources of heterogeneity such as different susceptibilities and likelihood of symptoms based on age, and also allows an easy parametrization of interventions such as school closures and/or lockdowns through the contact matrix. Note that the different categories need not be related to age, and could be used to represent other distinguishing properties such as behavior, physiology, employment, etc.

For each group $i$, let us define $S_i(t)$, $I_i(t)$ and $R_i(t)$ as the numbers of Susceptible, Infectious, and Recovered individuals in that group at time $t$. These structured models follow the SIR dynamics through a coupled system of ordinary differential equations written as
\begin{subequations}
\label{eq:SIR}
\begin{align}
    \dot{S}_i(t) &= -\beta\sigma_i \sum_{j}M_{ij}\frac{I_j}{N}S_i \\
    \dot{I}_i(t) &= \beta\sigma_i \sum_{j}M_{ij}\frac{I_j}{N}S_i - \gamma I_i \\
    \dot{R}_i(t) &= \gamma I_i
\end{align}
\end{subequations}
where $\beta$ is the baseline transmission rate of the disease, $\sigma_i$ is the group-dependent susceptibility of individuals, $M_{ij}$ is the average number of contacts someone in group $i$ has with people in group $j$ and through which that individual in group $i$ could become infected, and $\gamma$ is the recovery rate of the disease. Note that these equations are  equivalent to the model described in the Supplementary Materials of Ref.~\citep{zhang2020changes}.  They can be understood as setting a Poisson distribution of in-degree with average $z_{ij}^\mathrm{in} = \beta\sigma_i M_{ij}/\gamma$  coming from group $j$ for all individuals in group $i$. The model therefore corresponds to a mixture of directed Poisson processes, and it is unclear a priori how much we should expect these models to differ from the simple mass-action SIR.  Note that Eqs.~\eqref{eq:SIR} do not require $M_{ij}$ to be constrained by $M_{ji}$ in any way since it can be seen as quantifying the \textit{risk} that individuals in group $i$ catch the disease through their contacts with individuals in group $j$ (e.g. a musician playing along during a choir practice and who does not sing).  The matrix $\mathbf{M}$ whose elements are $\{M_{ij}\}$ therefore need not be symmetric, meaning that Eqs.~\eqref{eq:SIR} implicitly recognize that interactions between individuals can be directed.

We apply this framework to a multitype directed graph~\citep{allard2015general} (see Box~VI) where the type of a vertex corresponds to their age group and where directed edges allow us to consider heterogeneous susceptibility.  As shown in Fig.~\ref{fig:SIR}(a), the multitype directed graph approach exactly solves for the final epidemic size produced by the system presented in Eq.~(\ref{eq:SIR}).  The equations presented in Box~VI therefore provide a general solution for the epidemic size in structured SIR-like models.  Perhaps more importantly, this solution highlights how the solution of such systems is actually a simple mixture of Poisson processes and therefore a combination of Kermack-McKendrick-like terms.  As can be observed in Fig.~\ref{fig:SIR}(a), this type of heterogeneity based on realistic contact patterns does not translate into epidemic sizes that are very different from the ones predicted by the more simpler Kermack-McKendrick solution given by Eq.~\eqref{eq:KMK_solution}.

One intuition for this is that, since the sum of independent Poisson variables is Poisson (with the sum of their means), the total risk of individuals in these models, i.e. the number of incoming edges they have from all age groups, is Poisson-distributed.  Thus, while these models are structured, they result in a sum of terms like that in the original Kermack-McKendrick model, where $R_0$ is given by a weighted average (specifically the spectral radius of a matrix, see Box~VI).  As a result, mixtures of uncorrelated Poisson processes cannot truly embrace the full heterogeneity of risk and spread produced by biology, demographics, and human behavior.

\section{Conclusion\label{sec:conclusion}}
In this paper, we reviewed key results from random directed graphs to clarify the role of heterogeneity and correlations in risk and spread of an infectious disease. We noted that in directed graphs the friendship paradox plays different roles in contact tracing when following edges forward or backward: forward tracing leads to high risk individuals, and backward tracing leads to high-spread individuals. 

We then focused on identifying when the final epidemic size deviates from the classic Kermack-McKendrick solution. In particular, we showed that implementing heterogeneity in differential equation models by having a small number of types of individuals (representing, for example, age, occupation, susceptibility, and/or activity level) does not significantly affect epidemic size. Risk in these models is a mixture of Poisson distributions which individually correspond to Kermack-McKendrick-like contributions and therefore result in a similar epidemic size.
While contact patterns between groups may be important in assessing interventions (e.g. the effect of school closings) the epidemic size is not always as sensitive to these parameters as one might think.

The robustness of the Kermack-McKendrick result on directed graphs highlights the fact that the critical assumption behind their analysis of the SIR model is not mass-action mixing \textit{per se}, but instead an implicit assumption that risk is Poisson distributed and that ``risk'' and ``spread'' are independent. As we showed in Section~\ref{sec:examples}, this also implies that any model or simulation will fall back on the classic mass-action result if the secondary cases caused by an infectious individual are randomly selected from the population, no matter how heterogeneous the distribution of spread is.
When there is genuine heterogeneity in both risk and spread, they play dual roles in the early and late stages of an epidemic: the probability that an outbreak becomes an epidemic is driven primarily by superspreading events~\citep{althouse2020superspreading, lloyd-smith2005superspreading}, while the final epidemic size is driven primarily by high-risk individuals.

For emerging infectious diseases, especially those like COVID-19 where no proven pharmaceutical therapies nor prophylactics exist, our results highlight the critical need for detailed, contact tracing studies.  These studies should quantify both the incoming and outgoing risk of transmission and quantify the relative contribution of these risks during the epidemic.  Indeed, the efficacy of measures such as digital contact tracing~\citep{ferretti2020quantifying} and ``immune shielding"~\citep{weitz2020modeling} will depend on how risk and contact heterogeneity are structured across populations. Only through the application of models able to capture relevant heterogeneity, and the collection of empirical data on contact networks and risk, can the efficacy of such non-pharmaceutical interventions be accurately determined.

\section*{Acknowledgments}
A.A. acknowledges financial support from the Sentinelle Nord initiative of the Canada First Research Excellence Fund and from the Natural Sciences and Engineering Research Council of Canada (project 2019-05183). C.M. is supported by NSF Grant IIS-1838251. B.M.A. is supported by Bill and Melinda Gates through the Global Good Fund. S.V.S. is supported by startup funds provided by Northeastern University. L.H.-D. acknowledges support from the National Institutes of Health 1P20 GM125498-01 Centers of Biomedical Research Excellence Award.

\appendix
\section*{Mathematical case studies\label{sec:math}}

In this Appendix, we formally solve for the final epidemic size found in directed graphs representing different models of disease spread, including the examples used in Section~\ref{sec:examples}.

\subsection{Uncorrelated in- and out-degrees}

Let us first illustrate the formalism when in- and out-degree are independent. Denoting their respective PGFs by $\Gin(x)$ and $\Gout(y)$, Eqs.~\eqref{eq:G}, \eqref{eq:uout}, \eqref{eq:epidemic-prob}, \eqref{eq:uin} and \eqref{eq:epidemic-size} become
\begin{equation}
    \G(x, y)
        = \G(x,1) \,\G(1,y)
        = \Gin(x) \,\Gout(y)
\end{equation}
and
\begin{subequations}
\label{eq:solution_independence}
\begin{align}
    \uout & = \Gout(\uout) \ ;
    &
    \Sout & = 1 - \Gout(\uout) \ ; \\
    \uin & = \Gin(\uin) \ ;
    &
    \Sin & = 1 - \Gin(\uin) \label{eq:solution_independence_size} \ .
\end{align}
\end{subequations}
We conclude that the size of a macroscopic outbreak, $\Sin$, is solely dictated by the distribution of the in-degrees, i.e. by the distribution of the risk of individuals, while the probability for an outbreak to become macroscopic depends only on the distribution of spread.  Moreover, we see from Eq.~\eqref{eq:r-directed} that the number of secondary infections from patient zero and from secondary cases are equal, that is
\begin{align}
    R_0 = \frac{\langle \kin \kout \rangle}{\langle \kin \rangle}
        = \frac{\langle \kin \rangle \langle \kout \rangle}{\langle \kin \rangle}
        = \langle \kout \rangle
        = \langle \kin \rangle \ ,
\end{align}
and conclude therefore that there is no ``friendship paradox effect'' (from the \textit{excess} degree distribution) when in- and out-degree are not correlated.

\subsection{Derivation of Kermack-McKendrick\label{sec:kmk_derivation}}
In addition to being independent from the out-degrees, if we assume that the in-degrees are distributed according to a Poisson distribution, that is if $\Gin(x) = \mathrm{e}^{\langle \kin \rangle (x - 1)}$, Eq.~\eqref{eq:solution_independence_size} becomes
\begin{subequations}
\begin{align}
    \uin & = \mathrm{e}^{\langle \kin \rangle (\uin - 1)} \ ;
    &
    \Sin & = 1 - \mathrm{e}^{\langle \kin \rangle (\uin - 1)}
\end{align}
\end{subequations}
from which we obtain the results by Kermack-McKendrick for the size of an epidemic~\citep{kermack1927contribution, kermack1932contributions, kermack1933contributions}
\begin{align}
    \Sin = 1 - \mathrm{e}^{-R_0 \Sin } \ .
\end{align}
We conclude that the predictions for the size of a macroscopic outbreak will coincide with the Kermack-McKendrick solution given by Eq.~\eqref{eq:KMK_solution} if the in-degrees are not correlated with the out-degrees and if they are distributed according to a Poisson distribution (i.e. uniform risk).

To understand how uniform risk leads to a Poisson in-degree distribution, let us assume that each individual $i$ selects a finite number of other individuals, $k_\mathrm{out}^{(i)}$, that they could transmit to should they become infected.  This finite number can vary from one individual to another and is distributed according to the out-degree distribution.  In the limit of a large population size $n \to \infty$, the in-degree of individual $j$ due to individual $i$ is distributed according to $\mathrm{binomial}\big(k_\mathrm{out}^{(i)}, 1/(n-1)\big)$ whose PGF is $\left[ 1 + (x-1)/(n-1) \right]^{k_\mathrm{out}^{(i)}}$.  The PGF for the in-degree of individual $j$, obtained by considering the contribution of all $n-1$ other individuals, is
\begin{align} \label{eq:uniform_means_poisson}
    \G_\mathrm{in}^{(j)}(x)
        &  = \prod_{i \neq j} \left[ 1 + \frac{x-1}{n-1} \right]^{k_\mathrm{out}^{(i)}}
          \approx \prod_{i \neq j} \Bigg[ 1 + \frac{k_\mathrm{out}^{(i)}\ (x-1)}{n-1} \Bigg] \nonumber \\
        & \approx \Bigg[ 1 + \sum_{i\neq j}\frac{k_\mathrm{out}^{(i)}\ (x-1)}{n-1} \Bigg]
          \approx \Bigg[ 1 + \frac{(n-1)\ \langle k_\mathrm{out}^{(i)}\rangle\ (x-1)}{n-1} \Bigg] \nonumber \\
        & \approx \Bigg[ 1 + \frac{\langle k_\mathrm{out}^{(i)}\rangle\ (x-1)}{n-1} \Bigg]^{(n-1)}
          \approx \mathrm{e}^{\langle k_\mathrm{out}^{(i)}\rangle\ (x-1)}
\end{align}
where we used the binomial approximation twice, the definition of the exponential function and approximated $\langle k_\mathrm{out}^{(i)}\rangle \approx (n - 1)^{-1} \sum_{i\neq j} k_\mathrm{out}^{(i)}$.  Equation~\eqref{eq:uniform_means_poisson} tells us that the in-degree of every individual will be identically distributed in the limit of large population size, and that this in-degree distribution is a Poisson.

\subsection{Perfect correlation\label{sec:full_correlation}}
Let us now investigate the effect of a perfect correlation between individual risk and spread, which can be encoded as
\begin{equation}
    \G(x, y)
        = \sum_{i,j = 0}^{\infty} \Pin(i) \delta_{ij} x^i y^j
        = \sum_{i = 0}^{\infty} \Pin(i) (xy)^i
        \equiv \Gin(xy) \ ,
\end{equation}
where $\Pin(i)$ is the in-degree, or risk, distribution.  Note that perfect correlation does not imply that edges are reciprocal (i.e. edges run in both directions between vertices, which would be equivalent to undirected graphs).  Equations~\eqref{eq:uout}, \eqref{eq:epidemic-prob}, \eqref{eq:uin} and \eqref{eq:epidemic-size} become
\begin{subequations}
\label{eq:solution_full_correlation}
\begin{align}
    \uout & = \left[ \frac{1}{\langle \kin \rangle} \frac{\partial \Gin(xy)}{\partial x} \right]_{xy=\uout} \ ;
    &
    \Sout & = 1 - \Gin(\uout) \ ; \\
    \uin  & = \left[ \frac{1}{\langle \kin \rangle} \frac{\partial \Gin(xy)}{\partial y} \right]_{xy=\uin} \ ;
    &
    \Sin & = 1 - \Gin(\uin) \ . \label{eq:solution_full_correlation_Sin}
\end{align}
\end{subequations}
We conclude that in the presence of perfect correlation, the size and the probability of a macroscopic outbreak are equal.  Also, Eq.~\eqref{eq:r-directed} shows that
\begin{equation} \label{eq:R0_fully_correlated}
    R_0 = \frac{\langle \kin^2 \rangle}{\langle \kin \rangle}
\end{equation}
which differs from its undirected counterpart (i.e. $R_0 = \langle k (k - 1) \rangle / \langle k \rangle$)~\citep{newman2001random}.

As shown in Ref.~\citep{hebert-dufresne2020r0}, the undirected case, in which in- and out-degrees are perfectly correlated by construction, falls back on the result of Kermack-McKendrick when the degrees are distributed according to a Poisson distribution. However, this results does not extend to fully-correlated directed graphs.  Indeed, substituting $\Gin(x) = \mathrm{e}^{\langle \kin \rangle (x - 1)}$ in Eq.~\eqref{eq:solution_full_correlation_Sin} yields only two solutions $\uin=0,1$ for all $\langle \kin \rangle$. The probability for a macroscopic outbreak and its size are therefore equal to the fraction of vertices with nonzero in-/out-degree
\begin{align}
    \Sin & = \Sout = 1 - \mathrm{e}^{-\langle \kin \rangle} \ ,
\end{align}
and Eq.~\eqref{eq:R0_fully_correlated} yields $R_0 = \langle \kin \rangle + 1$, since a vertex that has been infected can always infect at least one other vertex.

\subsection{Arbitrary correlations\label{sec:arbitrary_correlations}}
To explore scenarios in-between the two previous limiting correlation cases, we consider the following PGF which allows to fix the marginal out-degree distribution and to interpolate between perfect correlation ($a = 1$) and independence ($a = 0$)
\begin{align} \label{eq:general_G}
    G(x,y) = [\Gout(xy)]^{a} [\Gin(x)\Gout(y) ]^{1-a} \ .
\end{align}
Note that in the case of perfect correlations, the in-degrees are distributed according to $\Gout$.  Equations~\eqref{eq:uout}, \eqref{eq:epidemic-prob}, \eqref{eq:uin} and \eqref{eq:epidemic-size} become
\begin{subequations}
\begin{align}
    \uout & = a \left[ \frac{1}{\langle \kin \rangle} \frac{\partial \Gout(xy)}{\partial x} \right]_{xy=\uout}
    + (1 - a) \Gout(\uout) \\
    \Sout & = 1 - \Gout(\uout) \\
    \uin & = a \left[ \frac{\Gin(\uin)}{\Gout(\uin)} \right]^{1-a} \left[ \frac{1}{\langle \kin \rangle} \frac{\partial \Gout(xy)}{\partial y} \right]_{xy=\uin}
    + (1 - a) \left[ \frac{\Gout(\uin)}{\Gin(\uin)} \right]^{a} \Gin(\uin) \label{eq:uin_correlations}\\
    \Sin & = 1 - [\Gout(\uin)]^{a} [\Gin(\uin)]^{1-a} \ .
\end{align}
\end{subequations}
From Eq.~\eqref{eq:r-directed}, we find that
\begin{align} \label{eq:R0_correlation}
    R_0 = (1 - a) \langle \kin \rangle + a \frac{\langle \kout^2 \rangle}{\langle \kin \rangle} \ .
\end{align}

\subsection{Correlated Poisson mixture model\label{sec:corr_poiss_mixt_model}}
\renewcommand{\uin}[1]{{u_{#1}^\mathrm{in}}}
We now combine the formalism of the previous section with the Poisson mixture model presented in Box~VI.  We consider a case in which there are 2 types of vertices, each type representing a fraction $w_1$ and $w_2$ of the graph, respectively.  We define $z_{ij}^\mathrm{in}$ as the average number of incoming edges vertices of type $j$ receive from vertices of type $j$, and $z_{ij}^\mathrm{out}$ as the average number of outgoing edges from vertices of type $i$ towards vertices of type $j$.  As in Box~VI, the parameters $\{w_i\}$, $\{z_{ij}^\mathrm{in}\}$ and $\{z_{ij}^\mathrm{out}\}$ are related via
\begin{equation}
  w_i z_{ij}^\mathrm{out} = w_j z_{ji}^\mathrm{in}
\end{equation}
for $i,j = 1,2$.  We assume that the number of incoming and outgoing edges between vertices is given by a Poisson mixture whose associated PGF take the following form
\begin{subequations}
\label{eq:corr_poiss_mist_G}
\begin{align}
    G_1(\bm{x},\bm{y})
        & = \exp \Big[       a    z_{11}^\mathrm{in}  (x_{11}y_{11} - 1) +    a    z_{12}^\mathrm{in}  (x_{12}y_{12} - 1) \nonumber \\
        &   \qquad \qquad (1 - a) z_{11}^\mathrm{in}  (x_{11}       - 1) + (1 - a) z_{12}^\mathrm{in}  (x_{12}       - 1) \nonumber \\
        &   \qquad \qquad (1 - a) z_{11}^\mathrm{out} (      y_{11} - 1) + (1 - a) z_{12}^\mathrm{out} (      y_{12} - 1) \Big]     \\
    G_2(\bm{x}, \bm{y})
        & = \exp \Big[            z_{21}^\mathrm{in}  (x_{21}       - 1) +         z_{22}^\mathrm{in}  (x_{22}       - 1) \nonumber \\
        &   \qquad \qquad         z_{21}^\mathrm{out} (      y_{21} - 1) +         z_{22}^\mathrm{out} (      y_{22} - 1) \Big]
\end{align}
\end{subequations}
where $\bm{x} = \big( x_{11}, x_{12}, x_{21}, x_{22} \big)$ and $\bm{y} = \big( y_{11}, y_{12}, y_{21}, y_{22} \big)$, and where $0 \leq a \leq 1$ controls the correlation between the in-degree and out-degree of vertices of type 1.  Perfect correlation is obtained when $a=1$, and $a=0$ corresponds to the uncorrelated case, similarly to the formalism presented in the previous section.

Inserting Eqs.~\eqref{eq:corr_poiss_mist_G} into the same PGF formalism as the one used to obtain the equations in Box~VI~\cite{allard2015general}, we obtain the following expression for the epidemic size
\begin{align}
    \Sin = 1 - \sum_{i=1}^2 w_i G_i(\bm{\uin{}}, \bm{1}) \ ,
\end{align}
where $\bm{\uin{}} = \big( \uin{11}, \uin{12}, \uin{21}, \uin{22} \big)$ is the smallest non-negative solution of the following system of equations
\begin{subequations}
\begin{align}
    \uin{11} & = (a \uin{11} + 1 - a) G_1(\bm{\uin{}}, \bm{1}) \\
    \uin{21} & = (a \uin{12} + 1 - a) G_1(\bm{\uin{}}, \bm{1}) \\
    \uin{12} & = \uin{22} = G_2(\bm{\uin{}}, \bm{1}) \ .
\end{align}
\end{subequations}

As for the Poisson mixture model described in Box~VI, the reproduction number, $R_0 = \rho(\mathbf{Z})$, is equal to the spectral radius of matrix $\mathbf{Z}$ which now takes the form
\begin{align}
    \mathbf{Z} =
        \begin{bmatrix}
          a + z_{11}^\mathrm{in} &     z_{12}^\mathrm{in} & 0                  & 0                  \\
          0                      & 0                      & z_{21}^\mathrm{in} & z_{22}^\mathrm{in} \\
              z_{11}^\mathrm{in} & a + z_{12}^\mathrm{in} & 0                  & 0                  \\
          0                      & 0                      & z_{21}^\mathrm{in} & z_{22}^\mathrm{in}
        \end{bmatrix} \ .
\end{align}

%


\begin{thebibliography}{50}%
\makeatletter
\providecommand \@ifxundefined [1]{%
 \@ifx{#1\undefined}
}%
\providecommand \@ifnum [1]{%
 \ifnum #1\expandafter \@firstoftwo
 \else \expandafter \@secondoftwo
 \fi
}%
\providecommand \@ifx [1]{%
 \ifx #1\expandafter \@firstoftwo
 \else \expandafter \@secondoftwo
 \fi
}%
\providecommand \natexlab [1]{#1}%
\providecommand \enquote  [1]{``#1''}%
\providecommand \bibnamefont  [1]{#1}%
\providecommand \bibfnamefont [1]{#1}%
\providecommand \citenamefont [1]{#1}%
\providecommand \href@noop [0]{\@secondoftwo}%
\providecommand \href [0]{\begingroup \@sanitize@url \@href}%
\providecommand \@href[1]{\@@startlink{#1}\@@href}%
\providecommand \@@href[1]{\endgroup#1\@@endlink}%
\providecommand \@sanitize@url [0]{\catcode `\\12\catcode `\$12\catcode
  `\&12\catcode `\#12\catcode `\^12\catcode `\_12\catcode `\%12\relax}%
\providecommand \@@startlink[1]{}%
\providecommand \@@endlink[0]{}%
\providecommand \url  [0]{\begingroup\@sanitize@url \@url }%
\providecommand \@url [1]{\endgroup\@href {#1}{\urlprefix }}%
\providecommand \urlprefix  [0]{URL }%
\providecommand \Eprint [0]{\href }%
\providecommand \doibase [0]{https://doi.org/}%
\providecommand \selectlanguage [0]{\@gobble}%
\providecommand \bibinfo  [0]{\@secondoftwo}%
\providecommand \bibfield  [0]{\@secondoftwo}%
\providecommand \translation [1]{[#1]}%
\providecommand \BibitemOpen [0]{}%
\providecommand \bibitemStop [0]{}%
\providecommand \bibitemNoStop [0]{.\EOS\space}%
\providecommand \EOS [0]{\spacefactor3000\relax}%
\providecommand \BibitemShut  [1]{\csname bibitem#1\endcsname}%
\let\auto@bib@innerbib\@empty
\bibitem [{\citenamefont {Kiss}\ \emph {et~al.}(2017)\citenamefont {Kiss},
  \citenamefont {Miller},\ and\ \citenamefont {Simon}}]{kiss2017mathematics}%
  \BibitemOpen
  \bibfield  {author} {\bibinfo {author} {\bibfnamefont {I.~Z.}\ \bibnamefont
  {Kiss}}, \bibinfo {author} {\bibfnamefont {J.~C.}\ \bibnamefont {Miller}},\
  and\ \bibinfo {author} {\bibfnamefont {P.~L.}\ \bibnamefont {Simon}},\ }\href
  {https://doi.org/10.1007/978-3-319-50806-1} {\emph {\bibinfo {title}
  {Mathematics of {{Epidemics}} on {{Networks}}: {{From Exact}} to
  {{Approximate Models}}}}}\ (\bibinfo  {publisher} {{Springer}},\ \bibinfo
  {year} {2017})\BibitemShut {NoStop}%
\bibitem [{\citenamefont {Newman}(2002)}]{newman2002spread}%
  \BibitemOpen
  \bibfield  {author} {\bibinfo {author} {\bibfnamefont {M.~E.~J.}\
  \bibnamefont {Newman}},\ }\bibfield  {title} {\bibinfo {title} {Spread of
  epidemic disease on networks},\ }\href
  {https://doi.org/10.1103/PhysRevE.66.016128} {\bibfield  {journal} {\bibinfo
  {journal} {Phys. Rev. E}\ }\textbf {\bibinfo {volume} {66}},\ \bibinfo
  {pages} {016128} (\bibinfo {year} {2002})}\BibitemShut {NoStop}%
\bibitem [{\citenamefont {Meyers}(2007)}]{meyers2007contact}%
  \BibitemOpen
  \bibfield  {author} {\bibinfo {author} {\bibfnamefont {L.~A.}\ \bibnamefont
  {Meyers}},\ }\bibfield  {title} {\bibinfo {title} {Contact network
  epidemiology: {{Bond}} percolation applied to infectious disease prediction
  and control},\ }\href {https://doi.org/10.1090/S0273-0979-06-01148-7}
  {\bibfield  {journal} {\bibinfo  {journal} {Bull. Am. Math. Soc.}\ }\textbf
  {\bibinfo {volume} {44}},\ \bibinfo {pages} {63} (\bibinfo {year}
  {2007})}\BibitemShut {NoStop}%
\bibitem [{\citenamefont {{H{\'e}bert-Dufresne}}\ \emph
  {et~al.}(2020)\citenamefont {{H{\'e}bert-Dufresne}}, \citenamefont
  {Althouse}, \citenamefont {Scarpino},\ and\ \citenamefont
  {Allard}}]{hebert-dufresne2020r0}%
  \BibitemOpen
  \bibfield  {author} {\bibinfo {author} {\bibfnamefont {L.}~\bibnamefont
  {{H{\'e}bert-Dufresne}}}, \bibinfo {author} {\bibfnamefont {B.~M.}\
  \bibnamefont {Althouse}}, \bibinfo {author} {\bibfnamefont {S.~V.}\
  \bibnamefont {Scarpino}},\ and\ \bibinfo {author} {\bibfnamefont
  {A.}~\bibnamefont {Allard}},\ }\bibfield  {title} {\bibinfo {title} {Beyond
  {{R0}}: heterogeneity in secondary infections and probabilistic epidemic
  forecasting},\ }\href {https://doi.org/10.1098/rsif.2020.0393} {\bibfield
  {journal} {\bibinfo  {journal} {J. R. Soc. Interface}\ }\textbf {\bibinfo
  {volume} {17}},\ \bibinfo {pages} {20200393} (\bibinfo {year}
  {2020})}\BibitemShut {NoStop}%
\bibitem [{\citenamefont {Kenah}\ and\ \citenamefont
  {Robins}(2007{\natexlab{a}})}]{kenah2007second}%
  \BibitemOpen
  \bibfield  {author} {\bibinfo {author} {\bibfnamefont {E.}~\bibnamefont
  {Kenah}}\ and\ \bibinfo {author} {\bibfnamefont {J.~M.}\ \bibnamefont
  {Robins}},\ }\bibfield  {title} {\bibinfo {title} {Second look at the spread
  of epidemics on networks},\ }\href
  {https://doi.org/10.1103/PhysRevE.76.036113} {\bibfield  {journal} {\bibinfo
  {journal} {Phys. Rev. E}\ }\textbf {\bibinfo {volume} {76}},\ \bibinfo
  {pages} {036113} (\bibinfo {year} {2007}{\natexlab{a}})}\BibitemShut
  {NoStop}%
\bibitem [{\citenamefont {Kenah}\ and\ \citenamefont
  {Robins}(2007{\natexlab{b}})}]{kenah2007networkbased}%
  \BibitemOpen
  \bibfield  {author} {\bibinfo {author} {\bibfnamefont {E.}~\bibnamefont
  {Kenah}}\ and\ \bibinfo {author} {\bibfnamefont {J.~M.}\ \bibnamefont
  {Robins}},\ }\bibfield  {title} {\bibinfo {title} {Network-based analysis of
  stochastic {{SIR}} epidemic models with random and proportionate mixing},\
  }\href {https://doi.org/10.1016/j.jtbi.2007.09.011} {\bibfield  {journal}
  {\bibinfo  {journal} {J. Theor. Biol.}\ }\textbf {\bibinfo {volume} {249}},\
  \bibinfo {pages} {706} (\bibinfo {year} {2007}{\natexlab{b}})}\BibitemShut
  {NoStop}%
\bibitem [{\citenamefont {Kenah}\ and\ \citenamefont
  {Miller}(2011)}]{kenah2011epidemic}%
  \BibitemOpen
  \bibfield  {author} {\bibinfo {author} {\bibfnamefont {E.}~\bibnamefont
  {Kenah}}\ and\ \bibinfo {author} {\bibfnamefont {J.~C.}\ \bibnamefont
  {Miller}},\ }\bibfield  {title} {\bibinfo {title} {Epidemic {{Percolation
  Networks}}, {{Epidemic Outcomes}}, and {{Interventions}}},\ }\href
  {https://doi.org/10.1155/2011/543520} {\bibfield  {journal} {\bibinfo
  {journal} {Interdiscip. Perspect. Infect. Dis.}\ }\textbf {\bibinfo {volume}
  {2011}},\ \bibinfo {pages} {543520} (\bibinfo {year} {2011})}\BibitemShut
  {NoStop}%
\bibitem [{\citenamefont {Miller}(2007)}]{miller2007epidemic}%
  \BibitemOpen
  \bibfield  {author} {\bibinfo {author} {\bibfnamefont {J.~C.}\ \bibnamefont
  {Miller}},\ }\bibfield  {title} {\bibinfo {title} {Epidemic size and
  probability in populations with heterogeneous infectivity and
  susceptibility},\ }\href {https://doi.org/10.1103/PhysRevE.76.010101}
  {\bibfield  {journal} {\bibinfo  {journal} {Phys. Rev. E}\ }\textbf {\bibinfo
  {volume} {76}},\ \bibinfo {pages} {010101} (\bibinfo {year}
  {2007})}\BibitemShut {NoStop}%
\bibitem [{\citenamefont {Anderson}\ and\ \citenamefont
  {May}(1992)}]{anderson1992infectious}%
  \BibitemOpen
  \bibfield  {author} {\bibinfo {author} {\bibfnamefont {R.~M.}\ \bibnamefont
  {Anderson}}\ and\ \bibinfo {author} {\bibfnamefont {R.~M.}\ \bibnamefont
  {May}},\ }\href
  {https://global.oup.com/academic/product/infectious-diseases-of-humans-9780198540403}
  {\emph {\bibinfo {title} {Infectious {{Diseases}} of {{Humans}}: {{Dynamics}}
  and {{Control}}}}}\ (\bibinfo  {publisher} {{Oxford University Press}},\
  \bibinfo {year} {1992})\BibitemShut {NoStop}%
\bibitem [{\citenamefont {Kermack}\ and\ \citenamefont
  {McKendrick}(1927)}]{kermack1927contribution}%
  \BibitemOpen
  \bibfield  {author} {\bibinfo {author} {\bibfnamefont {W.~O.}\ \bibnamefont
  {Kermack}}\ and\ \bibinfo {author} {\bibfnamefont {A.~G.}\ \bibnamefont
  {McKendrick}},\ }\bibfield  {title} {\bibinfo {title} {A {{Contribution}} to
  the {{Mathematical Theory}} of {{Epidemics}}},\ }\href
  {https://doi.org/10.1098/rspa.1927.0118} {\bibfield  {journal} {\bibinfo
  {journal} {Proc. R. Soc. A}\ }\textbf {\bibinfo {volume} {115}},\ \bibinfo
  {pages} {700} (\bibinfo {year} {1927})}\BibitemShut {NoStop}%
\bibitem [{\citenamefont {Kermack}\ and\ \citenamefont
  {McKendrick}(1932)}]{kermack1932contributions}%
  \BibitemOpen
  \bibfield  {author} {\bibinfo {author} {\bibfnamefont {W.~O.}\ \bibnamefont
  {Kermack}}\ and\ \bibinfo {author} {\bibfnamefont {A.~G.}\ \bibnamefont
  {McKendrick}},\ }\bibfield  {title} {\bibinfo {title} {Contributions to the
  {{Mathematical Theory}} of {{Epidemics}}. {{II}}. {{The Problem}} of
  {{Endemicity}}},\ }\href {https://doi.org/10.1098/rspa.1932.0171} {\bibfield
  {journal} {\bibinfo  {journal} {Proc. R. Soc. A}\ }\textbf {\bibinfo {volume}
  {138}},\ \bibinfo {pages} {55} (\bibinfo {year} {1932})}\BibitemShut
  {NoStop}%
\bibitem [{\citenamefont {Kermack}\ and\ \citenamefont
  {McKendrick}(1933)}]{kermack1933contributions}%
  \BibitemOpen
  \bibfield  {author} {\bibinfo {author} {\bibfnamefont {W.~O.}\ \bibnamefont
  {Kermack}}\ and\ \bibinfo {author} {\bibfnamefont {A.~G.}\ \bibnamefont
  {McKendrick}},\ }\bibfield  {title} {\bibinfo {title} {Contributions to the
  {{Mathematical Theory}} of {{Epidemics}}. {{III}}. {{Further Studies}} of the
  {{Problem}} of {{Endemicity}}},\ }\href
  {https://doi.org/10.1098/rspa.1933.0106} {\bibfield  {journal} {\bibinfo
  {journal} {Proc. R. Soc. A}\ }\textbf {\bibinfo {volume} {141}},\ \bibinfo
  {pages} {94} (\bibinfo {year} {1933})}\BibitemShut {NoStop}%
\bibitem [{\citenamefont {{Lloyd-Smith}}\ \emph {et~al.}(2005)\citenamefont
  {{Lloyd-Smith}}, \citenamefont {Schreiber}, \citenamefont {Kopp},\ and\
  \citenamefont {Getz}}]{lloyd-smith2005superspreading}%
  \BibitemOpen
  \bibfield  {author} {\bibinfo {author} {\bibfnamefont {J.~O.}\ \bibnamefont
  {{Lloyd-Smith}}}, \bibinfo {author} {\bibfnamefont {S.~J.}\ \bibnamefont
  {Schreiber}}, \bibinfo {author} {\bibfnamefont {P.~E.}\ \bibnamefont
  {Kopp}},\ and\ \bibinfo {author} {\bibfnamefont {W.~M.}\ \bibnamefont
  {Getz}},\ }\bibfield  {title} {\bibinfo {title} {Superspreading and the
  effect of individual variation on disease emergence},\ }\href
  {https://doi.org/10.1038/nature04153} {\bibfield  {journal} {\bibinfo
  {journal} {Nature}\ }\textbf {\bibinfo {volume} {438}},\ \bibinfo {pages}
  {355} (\bibinfo {year} {2005})}\BibitemShut {NoStop}%
\bibitem [{\citenamefont {Newman}(2018)}]{newman2018networks}%
  \BibitemOpen
  \bibfield  {author} {\bibinfo {author} {\bibfnamefont {M.~E.~J.}\
  \bibnamefont {Newman}},\ }\href
  {https://global.oup.com/academic/product/networks-9780198805090} {\emph
  {\bibinfo {title} {Networks}}}\ (\bibinfo  {publisher} {{Oxford University
  Press}},\ \bibinfo {year} {2018})\BibitemShut {NoStop}%
\bibitem [{\citenamefont {M{\"u}ller}\ \emph {et~al.}(2000)\citenamefont
  {M{\"u}ller}, \citenamefont {Kretzschmar},\ and\ \citenamefont
  {Dietz}}]{muller2000contact}%
  \BibitemOpen
  \bibfield  {author} {\bibinfo {author} {\bibfnamefont {J.}~\bibnamefont
  {M{\"u}ller}}, \bibinfo {author} {\bibfnamefont {M.}~\bibnamefont
  {Kretzschmar}},\ and\ \bibinfo {author} {\bibfnamefont {K.}~\bibnamefont
  {Dietz}},\ }\bibfield  {title} {\bibinfo {title} {Contact tracing in
  stochastic and deterministic epidemic models},\ }\href
  {https://doi.org/10.1016/S0025-5564(99)00061-9} {\bibfield  {journal}
  {\bibinfo  {journal} {Math. Biosci.}\ }\textbf {\bibinfo {volume} {164}},\
  \bibinfo {pages} {39} (\bibinfo {year} {2000})}\BibitemShut {NoStop}%
\bibitem [{\citenamefont {Eames}\ and\ \citenamefont
  {Keeling}(2003)}]{eames2003contact}%
  \BibitemOpen
  \bibfield  {author} {\bibinfo {author} {\bibfnamefont {K.~T.~D.}\
  \bibnamefont {Eames}}\ and\ \bibinfo {author} {\bibfnamefont {M.~J.}\
  \bibnamefont {Keeling}},\ }\bibfield  {title} {\bibinfo {title} {Contact
  tracing and disease control},\ }\href
  {https://doi.org/10.1098/rspb.2003.2554} {\bibfield  {journal} {\bibinfo
  {journal} {Proc. R. Soc. B}\ }\textbf {\bibinfo {volume} {270}},\ \bibinfo
  {pages} {2565} (\bibinfo {year} {2003})}\BibitemShut {NoStop}%
\bibitem [{\citenamefont {Kiss}\ \emph {et~al.}(2006)\citenamefont {Kiss},
  \citenamefont {Green},\ and\ \citenamefont {Kao}}]{kiss2006infectious}%
  \BibitemOpen
  \bibfield  {author} {\bibinfo {author} {\bibfnamefont {I.~Z.}\ \bibnamefont
  {Kiss}}, \bibinfo {author} {\bibfnamefont {D.~M.}\ \bibnamefont {Green}},\
  and\ \bibinfo {author} {\bibfnamefont {R.~R.}\ \bibnamefont {Kao}},\
  }\bibfield  {title} {\bibinfo {title} {Infectious disease control using
  contact tracing in random and scale-free networks},\ }\href
  {https://doi.org/10.1098/rsif.2005.0079} {\bibfield  {journal} {\bibinfo
  {journal} {J. R. Soc. Interface}\ }\textbf {\bibinfo {volume} {3}},\ \bibinfo
  {pages} {55} (\bibinfo {year} {2006})}\BibitemShut {NoStop}%
\bibitem [{\citenamefont {Hellewell}\ \emph {et~al.}(2020)\citenamefont
  {Hellewell}, \citenamefont {Abbott}, \citenamefont {Gimma}, \citenamefont
  {Bosse}, \citenamefont {Jarvis}, \citenamefont {Russell}, \citenamefont
  {Munday}, \citenamefont {Kucharski}, \citenamefont {Edmunds}, \citenamefont
  {Sun}, \citenamefont {Flasche}, \citenamefont {Quilty}, \citenamefont
  {Davies}, \citenamefont {Liu}, \citenamefont {Clifford}, \citenamefont
  {Klepac}, \citenamefont {Jit}, \citenamefont {Diamond}, \citenamefont
  {Gibbs}, \citenamefont {{van Zandvoort}}, \citenamefont {Funk},\ and\
  \citenamefont {Eggo}}]{hellewell2020feasibility}%
  \BibitemOpen
  \bibfield  {author} {\bibinfo {author} {\bibfnamefont {J.}~\bibnamefont
  {Hellewell}}, \bibinfo {author} {\bibfnamefont {S.}~\bibnamefont {Abbott}},
  \bibinfo {author} {\bibfnamefont {A.}~\bibnamefont {Gimma}}, \bibinfo
  {author} {\bibfnamefont {N.~I.}\ \bibnamefont {Bosse}}, \bibinfo {author}
  {\bibfnamefont {C.~I.}\ \bibnamefont {Jarvis}}, \bibinfo {author}
  {\bibfnamefont {T.~W.}\ \bibnamefont {Russell}}, \bibinfo {author}
  {\bibfnamefont {J.~D.}\ \bibnamefont {Munday}}, \bibinfo {author}
  {\bibfnamefont {A.~J.}\ \bibnamefont {Kucharski}}, \bibinfo {author}
  {\bibfnamefont {W.~J.}\ \bibnamefont {Edmunds}}, \bibinfo {author}
  {\bibfnamefont {F.}~\bibnamefont {Sun}}, \bibinfo {author} {\bibfnamefont
  {S.}~\bibnamefont {Flasche}}, \bibinfo {author} {\bibfnamefont {B.~J.}\
  \bibnamefont {Quilty}}, \bibinfo {author} {\bibfnamefont {N.}~\bibnamefont
  {Davies}}, \bibinfo {author} {\bibfnamefont {Y.}~\bibnamefont {Liu}},
  \bibinfo {author} {\bibfnamefont {S.}~\bibnamefont {Clifford}}, \bibinfo
  {author} {\bibfnamefont {P.}~\bibnamefont {Klepac}}, \bibinfo {author}
  {\bibfnamefont {M.}~\bibnamefont {Jit}}, \bibinfo {author} {\bibfnamefont
  {C.}~\bibnamefont {Diamond}}, \bibinfo {author} {\bibfnamefont
  {H.}~\bibnamefont {Gibbs}}, \bibinfo {author} {\bibfnamefont
  {K.}~\bibnamefont {{van Zandvoort}}}, \bibinfo {author} {\bibfnamefont
  {S.}~\bibnamefont {Funk}},\ and\ \bibinfo {author} {\bibfnamefont {R.~M.}\
  \bibnamefont {Eggo}},\ }\bibfield  {title} {\bibinfo {title} {Feasibility of
  controlling {{COVID}}-19 outbreaks by isolation of cases and contacts},\
  }\href {https://doi.org/10.1016/S2214-109X(20)30074-7} {\bibfield  {journal}
  {\bibinfo  {journal} {Lancet Glob. Health}\ }\textbf {\bibinfo {volume}
  {8}},\ \bibinfo {pages} {e488} (\bibinfo {year} {2020})}\BibitemShut
  {NoStop}%
\bibitem [{\citenamefont {Kojaku}\ \emph {et~al.}(2021)\citenamefont {Kojaku},
  \citenamefont {{H{\'e}bert-Dufresne}}, \citenamefont {Mones}, \citenamefont
  {Lehmann},\ and\ \citenamefont {Ahn}}]{kojaku2021effectiveness}%
  \BibitemOpen
  \bibfield  {author} {\bibinfo {author} {\bibfnamefont {S.}~\bibnamefont
  {Kojaku}}, \bibinfo {author} {\bibfnamefont {L.}~\bibnamefont
  {{H{\'e}bert-Dufresne}}}, \bibinfo {author} {\bibfnamefont {E.}~\bibnamefont
  {Mones}}, \bibinfo {author} {\bibfnamefont {S.}~\bibnamefont {Lehmann}},\
  and\ \bibinfo {author} {\bibfnamefont {Y.-Y.}\ \bibnamefont {Ahn}},\
  }\bibfield  {title} {\bibinfo {title} {The effectiveness of backward contact
  tracing in networks},\ }\href {https://doi.org/10.1038/s41567-021-01187-2}
  {\bibfield  {journal} {\bibinfo  {journal} {Nat. Phys.}\ }\textbf {\bibinfo
  {volume} {17}},\ \bibinfo {pages} {652} (\bibinfo {year} {2021})}\BibitemShut
  {NoStop}%
\bibitem [{\citenamefont {Meyers}\ \emph {et~al.}(2006)\citenamefont {Meyers},
  \citenamefont {Newman},\ and\ \citenamefont
  {Pourbohloul}}]{meyers2006predicting}%
  \BibitemOpen
  \bibfield  {author} {\bibinfo {author} {\bibfnamefont {L.~A.}\ \bibnamefont
  {Meyers}}, \bibinfo {author} {\bibfnamefont {M.~E.~J.}\ \bibnamefont
  {Newman}},\ and\ \bibinfo {author} {\bibfnamefont {B.}~\bibnamefont
  {Pourbohloul}},\ }\bibfield  {title} {\bibinfo {title} {Predicting epidemics
  on directed contact networks.},\ }\href
  {https://doi.org/10.1016/j.jtbi.2005.10.004} {\bibfield  {journal} {\bibinfo
  {journal} {J. Theor. Biol.}\ }\textbf {\bibinfo {volume} {240}},\ \bibinfo
  {pages} {400} (\bibinfo {year} {2006})}\BibitemShut {NoStop}%
\bibitem [{\citenamefont {Desai}\ and\ \citenamefont
  {Aronoff}(2020)}]{desai2020masks}%
  \BibitemOpen
  \bibfield  {author} {\bibinfo {author} {\bibfnamefont {A.~N.}\ \bibnamefont
  {Desai}}\ and\ \bibinfo {author} {\bibfnamefont {D.~M.}\ \bibnamefont
  {Aronoff}},\ }\bibfield  {title} {\bibinfo {title} {Masks and {{Coronavirus
  Disease}} 2019 ({{COVID}}-19)},\ }\href
  {https://doi.org/10.1001/jama.2020.6437} {\bibfield  {journal} {\bibinfo
  {journal} {JAMA}\ }\textbf {\bibinfo {volume} {323}},\ \bibinfo {pages}
  {2103} (\bibinfo {year} {2020})}\BibitemShut {NoStop}%
\bibitem [{\citenamefont {Greenhalgh}\ \emph {et~al.}(2020)\citenamefont
  {Greenhalgh}, \citenamefont {Schmid}, \citenamefont {Czypionka},
  \citenamefont {Bassler},\ and\ \citenamefont {Gruer}}]{greenhalgh2020face}%
  \BibitemOpen
  \bibfield  {author} {\bibinfo {author} {\bibfnamefont {T.}~\bibnamefont
  {Greenhalgh}}, \bibinfo {author} {\bibfnamefont {M.~B.}\ \bibnamefont
  {Schmid}}, \bibinfo {author} {\bibfnamefont {T.}~\bibnamefont {Czypionka}},
  \bibinfo {author} {\bibfnamefont {D.}~\bibnamefont {Bassler}},\ and\ \bibinfo
  {author} {\bibfnamefont {L.}~\bibnamefont {Gruer}},\ }\bibfield  {title}
  {\bibinfo {title} {Face masks for the public during the {{COVID}}-19
  crisis},\ }\href {https://doi.org/10.1136/bmj.m1435} {\bibfield  {journal}
  {\bibinfo  {journal} {BMJ}\ }\textbf {\bibinfo {volume} {369}},\ \bibinfo
  {pages} {m1435} (\bibinfo {year} {2020})}\BibitemShut {NoStop}%
\bibitem [{\citenamefont {Bansal}\ \emph {et~al.}(2006)\citenamefont {Bansal},
  \citenamefont {Pourbohloul},\ and\ \citenamefont
  {Meyers}}]{bansal2006comparative}%
  \BibitemOpen
  \bibfield  {author} {\bibinfo {author} {\bibfnamefont {S.}~\bibnamefont
  {Bansal}}, \bibinfo {author} {\bibfnamefont {B.}~\bibnamefont
  {Pourbohloul}},\ and\ \bibinfo {author} {\bibfnamefont {L.~A.}\ \bibnamefont
  {Meyers}},\ }\bibfield  {title} {\bibinfo {title} {A {{Comparative Analysis}}
  of {{Influenza Vaccination Programs}}},\ }\href
  {https://doi.org/10.1371/journal.pmed.0030387} {\bibfield  {journal}
  {\bibinfo  {journal} {PLOS Med.}\ }\textbf {\bibinfo {volume} {3}},\ \bibinfo
  {pages} {e387} (\bibinfo {year} {2006})}\BibitemShut {NoStop}%
\bibitem [{\citenamefont {Miller}(2008)}]{miller2008bounding}%
  \BibitemOpen
  \bibfield  {author} {\bibinfo {author} {\bibfnamefont {J.~C.}\ \bibnamefont
  {Miller}},\ }\bibfield  {title} {\bibinfo {title} {Bounding the {{Size}} and
  {{Probability}} of {{Epidemics}} on {{Networks}}},\ }\href
  {https://doi.org/10.1239/jap/1214950363} {\bibfield  {journal} {\bibinfo
  {journal} {J. Appl. Probab.}\ }\textbf {\bibinfo {volume} {45}},\ \bibinfo
  {pages} {498} (\bibinfo {year} {2008})}\BibitemShut {NoStop}%
\bibitem [{\citenamefont {Dorogovtsev}\ \emph {et~al.}(2001)\citenamefont
  {Dorogovtsev}, \citenamefont {Mendes},\ and\ \citenamefont
  {Samukhin}}]{dorogovtsev2001giant}%
  \BibitemOpen
  \bibfield  {author} {\bibinfo {author} {\bibfnamefont {S.~N.}\ \bibnamefont
  {Dorogovtsev}}, \bibinfo {author} {\bibfnamefont {J.~F.~F.}\ \bibnamefont
  {Mendes}},\ and\ \bibinfo {author} {\bibfnamefont {A.~N.}\ \bibnamefont
  {Samukhin}},\ }\bibfield  {title} {\bibinfo {title} {Giant strongly connected
  component of directed networks},\ }\href
  {https://doi.org/10.1103/PhysRevE.64.025101} {\bibfield  {journal} {\bibinfo
  {journal} {Phys. Rev. E}\ }\textbf {\bibinfo {volume} {64}},\ \bibinfo
  {pages} {025101} (\bibinfo {year} {2001})}\BibitemShut {NoStop}%
\bibitem [{\citenamefont {Bogu{\~n}{\'a}}\ and\ \citenamefont
  {Serrano}(2005)}]{boguna2005generalized}%
  \BibitemOpen
  \bibfield  {author} {\bibinfo {author} {\bibfnamefont {M.}~\bibnamefont
  {Bogu{\~n}{\'a}}}\ and\ \bibinfo {author} {\bibfnamefont {M.~{\'A}.}\
  \bibnamefont {Serrano}},\ }\bibfield  {title} {\bibinfo {title} {Generalized
  percolation in random directed networks},\ }\href
  {https://doi.org/10.1103/PhysRevE.72.016106} {\bibfield  {journal} {\bibinfo
  {journal} {Phys. Rev. E}\ }\textbf {\bibinfo {volume} {72}},\ \bibinfo
  {pages} {016106} (\bibinfo {year} {2005})}\BibitemShut {NoStop}%
\bibitem [{\citenamefont {Newman}\ \emph {et~al.}(2001)\citenamefont {Newman},
  \citenamefont {Strogatz},\ and\ \citenamefont {Watts}}]{newman2001random}%
  \BibitemOpen
  \bibfield  {author} {\bibinfo {author} {\bibfnamefont {M.~E.~J.}\
  \bibnamefont {Newman}}, \bibinfo {author} {\bibfnamefont {S.~H.}\
  \bibnamefont {Strogatz}},\ and\ \bibinfo {author} {\bibfnamefont {D.~J.}\
  \bibnamefont {Watts}},\ }\bibfield  {title} {\bibinfo {title} {Random graphs
  with arbitrary degree distributions and their applications},\ }\href
  {https://doi.org/10.1103/PhysRevE.64.026118} {\bibfield  {journal} {\bibinfo
  {journal} {Phys. Rev. E}\ }\textbf {\bibinfo {volume} {64}},\ \bibinfo
  {pages} {026118} (\bibinfo {year} {2001})}\BibitemShut {NoStop}%
\bibitem [{\citenamefont {Cooper}\ and\ \citenamefont
  {Frieze}(2004)}]{cooper2004size}%
  \BibitemOpen
  \bibfield  {author} {\bibinfo {author} {\bibfnamefont {C.}~\bibnamefont
  {Cooper}}\ and\ \bibinfo {author} {\bibfnamefont {A.}~\bibnamefont
  {Frieze}},\ }\bibfield  {title} {\bibinfo {title} {The {{Size}} of the
  {{Largest Strongly Connected Component}} of a {{Random Digraph}} with a
  {{Given Degree Sequence}}},\ }\href
  {https://doi.org/10.1017/S096354830400611X} {\bibfield  {journal} {\bibinfo
  {journal} {Comb. Probab. Comput.}\ }\textbf {\bibinfo {volume} {13}},\
  \bibinfo {pages} {319} (\bibinfo {year} {2004})}\BibitemShut {NoStop}%
\bibitem [{\citenamefont {Tim{\'a}r}\ \emph {et~al.}(2017)\citenamefont
  {Tim{\'a}r}, \citenamefont {Goltsev}, \citenamefont {Dorogovtsev},\ and\
  \citenamefont {Mendes}}]{timar2017mapping}%
  \BibitemOpen
  \bibfield  {author} {\bibinfo {author} {\bibfnamefont {G.}~\bibnamefont
  {Tim{\'a}r}}, \bibinfo {author} {\bibfnamefont {A.~V.}\ \bibnamefont
  {Goltsev}}, \bibinfo {author} {\bibfnamefont {S.~N.}\ \bibnamefont
  {Dorogovtsev}},\ and\ \bibinfo {author} {\bibfnamefont {J.~F.~F.}\
  \bibnamefont {Mendes}},\ }\bibfield  {title} {\bibinfo {title} {Mapping the
  {{Structure}} of {{Directed Networks}}: {{Beyond}} the {{Bow}}-{{Tie
  Diagram}}},\ }\href {https://doi.org/10.1103/PhysRevLett.118.078301}
  {\bibfield  {journal} {\bibinfo  {journal} {Phys. Rev. Lett.}\ }\textbf
  {\bibinfo {volume} {118}},\ \bibinfo {pages} {078301} (\bibinfo {year}
  {2017})}\BibitemShut {NoStop}%
\bibitem [{\citenamefont {Kuulasmaa}(1982)}]{kuulasmaa1982spatial}%
  \BibitemOpen
  \bibfield  {author} {\bibinfo {author} {\bibfnamefont {K.}~\bibnamefont
  {Kuulasmaa}},\ }\bibfield  {title} {\bibinfo {title} {The spatial general
  epidemic and locally dependent random graphs},\ }\href
  {https://doi.org/10.2307/3213827} {\bibfield  {journal} {\bibinfo  {journal}
  {J. Appl. Probab.}\ }\textbf {\bibinfo {volume} {19}},\ \bibinfo {pages}
  {745} (\bibinfo {year} {1982})}\BibitemShut {NoStop}%
\bibitem [{\citenamefont {Zhang}\ \emph {et~al.}(2020)\citenamefont {Zhang},
  \citenamefont {Litvinova}, \citenamefont {Liang}, \citenamefont {Wang},
  \citenamefont {Wang}, \citenamefont {Zhao}, \citenamefont {Wu}, \citenamefont
  {Merler}, \citenamefont {Viboud}, \citenamefont {Vespignani}, \citenamefont
  {Ajelli},\ and\ \citenamefont {Yu}}]{zhang2020changes}%
  \BibitemOpen
  \bibfield  {author} {\bibinfo {author} {\bibfnamefont {J.}~\bibnamefont
  {Zhang}}, \bibinfo {author} {\bibfnamefont {M.}~\bibnamefont {Litvinova}},
  \bibinfo {author} {\bibfnamefont {Y.}~\bibnamefont {Liang}}, \bibinfo
  {author} {\bibfnamefont {Y.}~\bibnamefont {Wang}}, \bibinfo {author}
  {\bibfnamefont {W.}~\bibnamefont {Wang}}, \bibinfo {author} {\bibfnamefont
  {S.}~\bibnamefont {Zhao}}, \bibinfo {author} {\bibfnamefont {Q.}~\bibnamefont
  {Wu}}, \bibinfo {author} {\bibfnamefont {S.}~\bibnamefont {Merler}}, \bibinfo
  {author} {\bibfnamefont {C.}~\bibnamefont {Viboud}}, \bibinfo {author}
  {\bibfnamefont {A.}~\bibnamefont {Vespignani}}, \bibinfo {author}
  {\bibfnamefont {M.}~\bibnamefont {Ajelli}},\ and\ \bibinfo {author}
  {\bibfnamefont {H.}~\bibnamefont {Yu}},\ }\bibfield  {title} {\bibinfo
  {title} {Changes in contact patterns shape the dynamics of the {{COVID}}-19
  outbreak in {{China}}},\ }\href {https://doi.org/10.1126/science.abb8001}
  {\bibfield  {journal} {\bibinfo  {journal} {Science}\ }\textbf {\bibinfo
  {volume} {368}},\ \bibinfo {pages} {1481} (\bibinfo {year}
  {2020})}\BibitemShut {NoStop}%
\bibitem [{\citenamefont {Allard}\ \emph {et~al.}(2017)\citenamefont {Allard},
  \citenamefont {Althouse}, \citenamefont {Scarpino},\ and\ \citenamefont
  {{H{\'e}bert-Dufresne}}}]{allard2017asymmetric}%
  \BibitemOpen
  \bibfield  {author} {\bibinfo {author} {\bibfnamefont {A.}~\bibnamefont
  {Allard}}, \bibinfo {author} {\bibfnamefont {B.~M.}\ \bibnamefont
  {Althouse}}, \bibinfo {author} {\bibfnamefont {S.~V.}\ \bibnamefont
  {Scarpino}},\ and\ \bibinfo {author} {\bibfnamefont {L.}~\bibnamefont
  {{H{\'e}bert-Dufresne}}},\ }\bibfield  {title} {\bibinfo {title} {Asymmetric
  percolation drives a double transition in sexual contact networks},\ }\href
  {https://doi.org/10.1073/pnas.1703073114} {\bibfield  {journal} {\bibinfo
  {journal} {Proc. Natl. Acad. Sci. U.S.A.}\ }\textbf {\bibinfo {volume}
  {114}},\ \bibinfo {pages} {8969} (\bibinfo {year} {2017})}\BibitemShut
  {NoStop}%
\bibitem [{\citenamefont {{H{\'e}bert-Dufresne}}\ and\ \citenamefont
  {Allard}(2019)}]{hebert-dufresne2019smeared}%
  \BibitemOpen
  \bibfield  {author} {\bibinfo {author} {\bibfnamefont {L.}~\bibnamefont
  {{H{\'e}bert-Dufresne}}}\ and\ \bibinfo {author} {\bibfnamefont
  {A.}~\bibnamefont {Allard}},\ }\bibfield  {title} {\bibinfo {title} {Smeared
  phase transitions in percolation on real complex networks},\ }\href
  {https://doi.org/10.1103/PhysRevResearch.1.013009} {\bibfield  {journal}
  {\bibinfo  {journal} {Phys. Rev. Research}\ }\textbf {\bibinfo {volume}
  {1}},\ \bibinfo {pages} {013009} (\bibinfo {year} {2019})}\BibitemShut
  {NoStop}%
\bibitem [{\citenamefont {Frisch}\ and\ \citenamefont
  {Hammersley}(1963)}]{frisch1963percolation}%
  \BibitemOpen
  \bibfield  {author} {\bibinfo {author} {\bibfnamefont {H.~L.}\ \bibnamefont
  {Frisch}}\ and\ \bibinfo {author} {\bibfnamefont {J.~M.}\ \bibnamefont
  {Hammersley}},\ }\bibfield  {title} {\bibinfo {title} {Percolation
  {{Processes}} and {{Related Topics}}},\ }\href
  {https://doi.org/10.1137/0111066} {\bibfield  {journal} {\bibinfo  {journal}
  {J. Soc. Ind. Appl. Math.}\ }\textbf {\bibinfo {volume} {11}},\ \bibinfo
  {pages} {894} (\bibinfo {year} {1963})}\BibitemShut {NoStop}%
\bibitem [{\citenamefont {Dietz}(1967)}]{dietz1967epidemics}%
  \BibitemOpen
  \bibfield  {author} {\bibinfo {author} {\bibfnamefont {K.}~\bibnamefont
  {Dietz}},\ }\bibfield  {title} {\bibinfo {title} {Epidemics and {{Rumours}}:
  {{A Survey}}},\ }\href {https://doi.org/10.2307/2982521} {\bibfield
  {journal} {\bibinfo  {journal} {J. R. Stat. Soc. A}\ }\textbf {\bibinfo
  {volume} {130}},\ \bibinfo {pages} {505} (\bibinfo {year}
  {1967})}\BibitemShut {NoStop}%
\bibitem [{\citenamefont {Mollison}(1977)}]{mollison1977spatial}%
  \BibitemOpen
  \bibfield  {author} {\bibinfo {author} {\bibfnamefont {D.}~\bibnamefont
  {Mollison}},\ }\bibfield  {title} {\bibinfo {title} {Spatial {{Contact
  Models}} for {{Ecological}} and {{Epidemic Spread}}},\ }\href
  {https://doi.org/10.1111/j.2517-6161.1977.tb01627.x} {\bibfield  {journal}
  {\bibinfo  {journal} {J. R. Stat. Soc. B}\ }\textbf {\bibinfo {volume}
  {39}},\ \bibinfo {pages} {283} (\bibinfo {year} {1977})}\BibitemShut
  {NoStop}%
\bibitem [{\citenamefont {Grassberger}(1983)}]{grassberger1983critical}%
  \BibitemOpen
  \bibfield  {author} {\bibinfo {author} {\bibfnamefont {P.}~\bibnamefont
  {Grassberger}},\ }\bibfield  {title} {\bibinfo {title} {On the critical
  behavior of the general epidemic process and dynamical percolation},\ }\href
  {https://doi.org/10.1016/0025-5564(82)90036-0} {\bibfield  {journal}
  {\bibinfo  {journal} {Math. Biosci.}\ }\textbf {\bibinfo {volume} {63}},\
  \bibinfo {pages} {157} (\bibinfo {year} {1983})}\BibitemShut {NoStop}%
\bibitem [{\citenamefont {Cardy}\ and\ \citenamefont
  {Grassberger}(1985)}]{cardy1985epidemic}%
  \BibitemOpen
  \bibfield  {author} {\bibinfo {author} {\bibfnamefont {J.~L.}\ \bibnamefont
  {Cardy}}\ and\ \bibinfo {author} {\bibfnamefont {P.}~\bibnamefont
  {Grassberger}},\ }\bibfield  {title} {\bibinfo {title} {Epidemic models and
  percolation},\ }\href {https://doi.org/10.1088/0305-4470/18/6/001} {\bibfield
   {journal} {\bibinfo  {journal} {J. Phys. A}\ }\textbf {\bibinfo {volume}
  {18}},\ \bibinfo {pages} {L267} (\bibinfo {year} {1985})}\BibitemShut
  {NoStop}%
\bibitem [{\citenamefont {Sander}\ \emph {et~al.}(2002)\citenamefont {Sander},
  \citenamefont {Warren}, \citenamefont {Sokolov}, \citenamefont {Simon},\ and\
  \citenamefont {Koopman}}]{sander2002percolation}%
  \BibitemOpen
  \bibfield  {author} {\bibinfo {author} {\bibfnamefont {L.~M.}\ \bibnamefont
  {Sander}}, \bibinfo {author} {\bibfnamefont {C.~P.}\ \bibnamefont {Warren}},
  \bibinfo {author} {\bibfnamefont {I.~M.}\ \bibnamefont {Sokolov}}, \bibinfo
  {author} {\bibfnamefont {C.}~\bibnamefont {Simon}},\ and\ \bibinfo {author}
  {\bibfnamefont {J.}~\bibnamefont {Koopman}},\ }\bibfield  {title} {\bibinfo
  {title} {Percolation on heterogeneous networks as a model for epidemics},\
  }\href {https://doi.org/10.1016/S0025-5564(02)00117-7} {\bibfield  {journal}
  {\bibinfo  {journal} {Math. Biosci.}\ }\textbf {\bibinfo {volume} {180}},\
  \bibinfo {pages} {293} (\bibinfo {year} {2002})}\BibitemShut {NoStop}%
\bibitem [{\citenamefont {Meyers}\ \emph {et~al.}(2005)\citenamefont {Meyers},
  \citenamefont {Pourbohloul}, \citenamefont {Newman}, \citenamefont
  {Skowronski},\ and\ \citenamefont {Brunham}}]{meyers2005network}%
  \BibitemOpen
  \bibfield  {author} {\bibinfo {author} {\bibfnamefont {L.~A.}\ \bibnamefont
  {Meyers}}, \bibinfo {author} {\bibfnamefont {B.}~\bibnamefont {Pourbohloul}},
  \bibinfo {author} {\bibfnamefont {M.~E.~J.}\ \bibnamefont {Newman}}, \bibinfo
  {author} {\bibfnamefont {D.~M.}\ \bibnamefont {Skowronski}},\ and\ \bibinfo
  {author} {\bibfnamefont {R.~C.}\ \bibnamefont {Brunham}},\ }\bibfield
  {title} {\bibinfo {title} {Network theory and {{SARS}}: {{Predicting}}
  outbreak diversity},\ }\href {https://doi.org/10.1016/j.jtbi.2004.07.026}
  {\bibfield  {journal} {\bibinfo  {journal} {J. Theor. Biol.}\ }\textbf
  {\bibinfo {volume} {232}},\ \bibinfo {pages} {71} (\bibinfo {year}
  {2005})}\BibitemShut {NoStop}%
\bibitem [{\citenamefont {Funk}\ and\ \citenamefont
  {Jansen}(2010)}]{funk2010interacting}%
  \BibitemOpen
  \bibfield  {author} {\bibinfo {author} {\bibfnamefont {S.}~\bibnamefont
  {Funk}}\ and\ \bibinfo {author} {\bibfnamefont {V.~A.~A.}\ \bibnamefont
  {Jansen}},\ }\bibfield  {title} {\bibinfo {title} {Interacting epidemics on
  overlay networks},\ }\href {https://doi.org/10.1103/PhysRevE.81.036118}
  {\bibfield  {journal} {\bibinfo  {journal} {Phys. Rev. E}\ }\textbf {\bibinfo
  {volume} {81}},\ \bibinfo {pages} {036118} (\bibinfo {year}
  {2010})}\BibitemShut {NoStop}%
\bibitem [{\citenamefont {Dimitrov}\ and\ \citenamefont
  {Meyers}(2010)}]{dimitrov2010mathematical}%
  \BibitemOpen
  \bibfield  {author} {\bibinfo {author} {\bibfnamefont {N.~B.}\ \bibnamefont
  {Dimitrov}}\ and\ \bibinfo {author} {\bibfnamefont {L.~A.}\ \bibnamefont
  {Meyers}},\ }\bibfield  {title} {\bibinfo {title} {Mathematical
  {{Approaches}} to {{Infectious Disease Prediction}} and {{Control}}},\ }in\
  \href {https://doi.org/10.1287/educ.1100.0075} {\emph {\bibinfo {booktitle}
  {Risk and {{Optimization}} in an {{Uncertain World}}}}},\ \bibinfo {series
  and number} {{{INFORMS TutORials}} in {{Operations Research}}}\ (\bibinfo
  {year} {2010})\ pp.\ \bibinfo {pages} {1--25}\BibitemShut {NoStop}%
\bibitem [{\citenamefont {Allard}\ \emph {et~al.}(2009)\citenamefont {Allard},
  \citenamefont {No{\"e}l}, \citenamefont {Dub{\'e}},\ and\ \citenamefont
  {Pourbohloul}}]{allard2009heterogeneous}%
  \BibitemOpen
  \bibfield  {author} {\bibinfo {author} {\bibfnamefont {A.}~\bibnamefont
  {Allard}}, \bibinfo {author} {\bibfnamefont {P.-A.}\ \bibnamefont
  {No{\"e}l}}, \bibinfo {author} {\bibfnamefont {L.~J.}\ \bibnamefont
  {Dub{\'e}}},\ and\ \bibinfo {author} {\bibfnamefont {B.}~\bibnamefont
  {Pourbohloul}},\ }\bibfield  {title} {\bibinfo {title} {Heterogeneous bond
  percolation on multitype networks with an application to epidemic dynamics},\
  }\href {https://doi.org/10.1103/PhysRevE.79.036113} {\bibfield  {journal}
  {\bibinfo  {journal} {Phys. Rev. E}\ }\textbf {\bibinfo {volume} {79}},\
  \bibinfo {pages} {036113} (\bibinfo {year} {2009})}\BibitemShut {NoStop}%
\bibitem [{\citenamefont {Allard}\ \emph {et~al.}(2015)\citenamefont {Allard},
  \citenamefont {{H{\'e}bert-Dufresne}}, \citenamefont {Young},\ and\
  \citenamefont {Dub{\'e}}}]{allard2015general}%
  \BibitemOpen
  \bibfield  {author} {\bibinfo {author} {\bibfnamefont {A.}~\bibnamefont
  {Allard}}, \bibinfo {author} {\bibfnamefont {L.}~\bibnamefont
  {{H{\'e}bert-Dufresne}}}, \bibinfo {author} {\bibfnamefont {J.-G.}\
  \bibnamefont {Young}},\ and\ \bibinfo {author} {\bibfnamefont {L.~J.}\
  \bibnamefont {Dub{\'e}}},\ }\bibfield  {title} {\bibinfo {title} {General and
  exact approach to percolation on random graphs},\ }\href
  {https://doi.org/10.1103/PhysRevE.92.062807} {\bibfield  {journal} {\bibinfo
  {journal} {Phys. Rev. E}\ }\textbf {\bibinfo {volume} {92}},\ \bibinfo
  {pages} {062807} (\bibinfo {year} {2015})}\BibitemShut {NoStop}%
\bibitem [{\citenamefont {Artime}\ and\ \citenamefont
  {De~Domenico}(2021)}]{artime2021percolation}%
  \BibitemOpen
  \bibfield  {author} {\bibinfo {author} {\bibfnamefont {O.}~\bibnamefont
  {Artime}}\ and\ \bibinfo {author} {\bibfnamefont {M.}~\bibnamefont
  {De~Domenico}},\ }\bibfield  {title} {\bibinfo {title} {Percolation on
  feature-enriched interconnected systems},\ }\href
  {https://doi.org/10.1038/s41467-021-22721-z} {\bibfield  {journal} {\bibinfo
  {journal} {Nat. Commun.}\ }\textbf {\bibinfo {volume} {12}},\ \bibinfo
  {pages} {2478} (\bibinfo {year} {2021})}\BibitemShut {NoStop}%
\bibitem [{\citenamefont {Fumanelli}\ \emph {et~al.}(2012)\citenamefont
  {Fumanelli}, \citenamefont {Ajelli}, \citenamefont {Manfredi}, \citenamefont
  {Vespignani},\ and\ \citenamefont {Merler}}]{fumanelli2012inferring}%
  \BibitemOpen
  \bibfield  {author} {\bibinfo {author} {\bibfnamefont {L.}~\bibnamefont
  {Fumanelli}}, \bibinfo {author} {\bibfnamefont {M.}~\bibnamefont {Ajelli}},
  \bibinfo {author} {\bibfnamefont {P.}~\bibnamefont {Manfredi}}, \bibinfo
  {author} {\bibfnamefont {A.}~\bibnamefont {Vespignani}},\ and\ \bibinfo
  {author} {\bibfnamefont {S.}~\bibnamefont {Merler}},\ }\bibfield  {title}
  {\bibinfo {title} {Inferring the {{Structure}} of {{Social Contacts}} from
  {{Demographic Data}} in the {{Analysis}} of {{Infectious Diseases Spread}}},\
  }\href {https://doi.org/10.1371/journal.pcbi.1002673} {\bibfield  {journal}
  {\bibinfo  {journal} {PLOS Comput. Biol.}\ }\textbf {\bibinfo {volume} {8}},\
  \bibinfo {pages} {e1002673} (\bibinfo {year} {2012})}\BibitemShut {NoStop}%
\bibitem [{\citenamefont {Mistry}\ \emph {et~al.}(2021)\citenamefont {Mistry},
  \citenamefont {Litvinova}, \citenamefont {{Pastore y Piontti}}, \citenamefont
  {Chinazzi}, \citenamefont {Fumanelli}, \citenamefont {Gomes}, \citenamefont
  {Haque}, \citenamefont {Liu}, \citenamefont {Mu}, \citenamefont {Xiong},
  \citenamefont {Halloran}, \citenamefont {Longini}, \citenamefont {Merler},
  \citenamefont {Ajelli},\ and\ \citenamefont
  {Vespignani}}]{mistry2021inferring}%
  \BibitemOpen
  \bibfield  {author} {\bibinfo {author} {\bibfnamefont {D.}~\bibnamefont
  {Mistry}}, \bibinfo {author} {\bibfnamefont {M.}~\bibnamefont {Litvinova}},
  \bibinfo {author} {\bibfnamefont {A.}~\bibnamefont {{Pastore y Piontti}}},
  \bibinfo {author} {\bibfnamefont {M.}~\bibnamefont {Chinazzi}}, \bibinfo
  {author} {\bibfnamefont {L.}~\bibnamefont {Fumanelli}}, \bibinfo {author}
  {\bibfnamefont {M.~F.~C.}\ \bibnamefont {Gomes}}, \bibinfo {author}
  {\bibfnamefont {S.~A.}\ \bibnamefont {Haque}}, \bibinfo {author}
  {\bibfnamefont {Q.-H.}\ \bibnamefont {Liu}}, \bibinfo {author} {\bibfnamefont
  {K.}~\bibnamefont {Mu}}, \bibinfo {author} {\bibfnamefont {X.}~\bibnamefont
  {Xiong}}, \bibinfo {author} {\bibfnamefont {M.~E.}\ \bibnamefont {Halloran}},
  \bibinfo {author} {\bibfnamefont {I.~M.}\ \bibnamefont {Longini}}, \bibinfo
  {author} {\bibfnamefont {S.}~\bibnamefont {Merler}}, \bibinfo {author}
  {\bibfnamefont {M.}~\bibnamefont {Ajelli}},\ and\ \bibinfo {author}
  {\bibfnamefont {A.}~\bibnamefont {Vespignani}},\ }\bibfield  {title}
  {\bibinfo {title} {Inferring high-resolution human mixing patterns for
  disease modeling},\ }\href {https://doi.org/10.1038/s41467-020-20544-y}
  {\bibfield  {journal} {\bibinfo  {journal} {Nat. Commun.}\ }\textbf {\bibinfo
  {volume} {12}},\ \bibinfo {pages} {323} (\bibinfo {year} {2021})}\BibitemShut
  {NoStop}%
\bibitem [{\citenamefont {Althouse}\ \emph {et~al.}(2020)\citenamefont
  {Althouse}, \citenamefont {Wenger}, \citenamefont {Miller}, \citenamefont
  {Scarpino}, \citenamefont {Allard}, \citenamefont {{H{\'e}bert-Dufresne}},\
  and\ \citenamefont {Hu}}]{althouse2020superspreading}%
  \BibitemOpen
  \bibfield  {author} {\bibinfo {author} {\bibfnamefont {B.~M.}\ \bibnamefont
  {Althouse}}, \bibinfo {author} {\bibfnamefont {E.~A.}\ \bibnamefont
  {Wenger}}, \bibinfo {author} {\bibfnamefont {J.~C.}\ \bibnamefont {Miller}},
  \bibinfo {author} {\bibfnamefont {S.~V.}\ \bibnamefont {Scarpino}}, \bibinfo
  {author} {\bibfnamefont {A.}~\bibnamefont {Allard}}, \bibinfo {author}
  {\bibfnamefont {L.}~\bibnamefont {{H{\'e}bert-Dufresne}}},\ and\ \bibinfo
  {author} {\bibfnamefont {H.}~\bibnamefont {Hu}},\ }\bibfield  {title}
  {\bibinfo {title} {Superspreading events in the transmission dynamics of
  {{SARS}}-{{CoV}}-2: {{Opportunities}} for interventions and control},\ }\href
  {https://doi.org/10.1371/journal.pbio.3000897} {\bibfield  {journal}
  {\bibinfo  {journal} {PLOS Biol.}\ }\textbf {\bibinfo {volume} {18}},\
  \bibinfo {pages} {e3000897} (\bibinfo {year} {2020})}\BibitemShut {NoStop}%
\bibitem [{\citenamefont {Ferretti}\ \emph {et~al.}(2020)\citenamefont
  {Ferretti}, \citenamefont {Wymant}, \citenamefont {Kendall}, \citenamefont
  {Zhao}, \citenamefont {Nurtay}, \citenamefont {{Abeler-D{\"o}rner}},
  \citenamefont {Parker}, \citenamefont {Bonsall},\ and\ \citenamefont
  {Fraser}}]{ferretti2020quantifying}%
  \BibitemOpen
  \bibfield  {author} {\bibinfo {author} {\bibfnamefont {L.}~\bibnamefont
  {Ferretti}}, \bibinfo {author} {\bibfnamefont {C.}~\bibnamefont {Wymant}},
  \bibinfo {author} {\bibfnamefont {M.}~\bibnamefont {Kendall}}, \bibinfo
  {author} {\bibfnamefont {L.}~\bibnamefont {Zhao}}, \bibinfo {author}
  {\bibfnamefont {A.}~\bibnamefont {Nurtay}}, \bibinfo {author} {\bibfnamefont
  {L.}~\bibnamefont {{Abeler-D{\"o}rner}}}, \bibinfo {author} {\bibfnamefont
  {M.}~\bibnamefont {Parker}}, \bibinfo {author} {\bibfnamefont
  {D.}~\bibnamefont {Bonsall}},\ and\ \bibinfo {author} {\bibfnamefont
  {C.}~\bibnamefont {Fraser}},\ }\bibfield  {title} {\bibinfo {title}
  {Quantifying {{SARS}}-{{CoV}}-2 transmission suggests epidemic control with
  digital contact tracing},\ }\href {https://doi.org/10.1126/science.abb6936}
  {\bibfield  {journal} {\bibinfo  {journal} {Science}\ }\textbf {\bibinfo
  {volume} {368}},\ \bibinfo {pages} {eabb6936} (\bibinfo {year}
  {2020})}\BibitemShut {NoStop}%
\bibitem [{\citenamefont {Weitz}\ \emph {et~al.}(2020)\citenamefont {Weitz},
  \citenamefont {Beckett}, \citenamefont {Coenen}, \citenamefont {Demory},
  \citenamefont {{Dominguez-Mirazo}}, \citenamefont {Dushoff}, \citenamefont
  {Leung}, \citenamefont {Li}, \citenamefont {M{\u a}g{\u a}lie}, \citenamefont
  {Park}, \citenamefont {{Rodriguez-Gonzalez}}, \citenamefont {Shivam},\ and\
  \citenamefont {Zhao}}]{weitz2020modeling}%
  \BibitemOpen
  \bibfield  {author} {\bibinfo {author} {\bibfnamefont {J.~S.}\ \bibnamefont
  {Weitz}}, \bibinfo {author} {\bibfnamefont {S.~J.}\ \bibnamefont {Beckett}},
  \bibinfo {author} {\bibfnamefont {A.~R.}\ \bibnamefont {Coenen}}, \bibinfo
  {author} {\bibfnamefont {D.}~\bibnamefont {Demory}}, \bibinfo {author}
  {\bibfnamefont {M.}~\bibnamefont {{Dominguez-Mirazo}}}, \bibinfo {author}
  {\bibfnamefont {J.}~\bibnamefont {Dushoff}}, \bibinfo {author} {\bibfnamefont
  {C.-Y.}\ \bibnamefont {Leung}}, \bibinfo {author} {\bibfnamefont
  {G.}~\bibnamefont {Li}}, \bibinfo {author} {\bibfnamefont {A.}~\bibnamefont
  {M{\u a}g{\u a}lie}}, \bibinfo {author} {\bibfnamefont {S.~W.}\ \bibnamefont
  {Park}}, \bibinfo {author} {\bibfnamefont {R.}~\bibnamefont
  {{Rodriguez-Gonzalez}}}, \bibinfo {author} {\bibfnamefont {S.}~\bibnamefont
  {Shivam}},\ and\ \bibinfo {author} {\bibfnamefont {C.~Y.}\ \bibnamefont
  {Zhao}},\ }\bibfield  {title} {\bibinfo {title} {Modeling shield immunity to
  reduce {{COVID}}-19 epidemic spread},\ }\href
  {https://doi.org/10.1038/s41591-020-0895-3} {\bibfield  {journal} {\bibinfo
  {journal} {Nat. Med.}\ }\textbf {\bibinfo {volume} {26}},\ \bibinfo {pages}
  {849} (\bibinfo {year} {2020})}\BibitemShut {NoStop}%
\end{thebibliography}
\end{document}